\documentclass[twocolumn]{aastex631}

\usepackage{amsmath}
\usepackage{booktabs}
\usepackage{enumerate}
\usepackage{tablefootnote}
\usepackage{makecell}
\usepackage{hyperref}

\begin{document}
\title{Identifying Chemically Coherent Stellar Families in the Galactic High-$\alpha$ Disk with Low-Resolution Spectroscopy}

\author[0000-0002-9332-2871]{Tianmin Wu}
\affiliation{CAS Key Laboratory of Optical Astronomy, National Astronomical Observatories, Chinese Academy of Sciences, Beijing 100101, People’s Republic of China}
\affiliation{School of Astronomy and Space Science, University of Chinese Academy of Sciences, Beijing 100049, People’s Republic of China}

\author[0000-0002-5818-8769]{Maosheng Xiang}\email{msxiang@nao.cas.cn}
\affiliation{CAS Key Laboratory of Optical Astronomy, National Astronomical Observatories, Chinese Academy of Sciences, Beijing 100101, People’s Republic of China}
\affiliation{Institute for Frontiers in Astronomy and Astrophysics, Beijing Normal University, Beijing 102206, People’s Republic of China}

\author{Jianrong Shi}
\affiliation{CAS Key Laboratory of Optical Astronomy, National Astronomical Observatories, Chinese Academy of Sciences, Beijing 100101, People’s Republic of China}
\affiliation{School of Astronomy and Space Science, University of Chinese Academy of Sciences, Beijing 100049, People’s Republic of China}

\author{Tiancheng Sun}
\affiliation{CAS Key Laboratory of Optical Astronomy, National Astronomical Observatories, Chinese Academy of Sciences, Beijing 100101, People’s Republic of China}

\author{Meng Zhang}
\affiliation{CAS Key Laboratory of Optical Astronomy, National Astronomical Observatories, Chinese Academy of Sciences, Beijing 100101, People’s Republic of China}
\affiliation{Max-Planck-Institut f\"ur Astronomie, K\"onigstuhl 17, D-69117 Heidelberg, Germany}



\begin{abstract}

Identification of chemical substructures in the Galactic high-$\alpha$ disk provides valuable insight into its formation history. In this work, we demonstrate the capability of low-resolution spectroscopy for this purpose by applying DBSCAN clustering to 87,275 red giants in the high-$\alpha$ disk using 18 elemental abundances derived from LAMOST low-resolution spectra ($R\lesssim1800$). We assess the effectiveness of this approach by comparing the clustering results from the low-resolution data with those obtained by applying the same procedure to high-resolution survey abundances and to mock catalogs. These comparisons suggest that, despite the relatively large uncertainties in individual elemental abundances derived from low-resolution spectra, the astrophysical coherence among different elements provides sufficient information to identify chemically coherent families that may share similar enrichment histories without necessarily sharing a common birth environment. After scanning the parameter space of the DBSCAN procedure and merging the results from different parameter combinations, we identify 133 candidate chemical families containing 4,642 unique stars (5.3\% of the high-$\alpha$ disk sample). These families exhibit diverse multi-element abundance patterns, with particularly prominent differences in neutron-capture elements at nearly identical [Fe/H] values. These families therefore offer new constraints on the diversity of chemical enrichment pathways and on the assembly history of the high-$\alpha$ disk. 
\end{abstract}


\section{Introduction} \label{sec:intro}
Our Galaxy's disk hosts a high-$\alpha$ component that is clearly distinct from the low-$\alpha$ population in the stellar [$\alpha$/Fe]--[Fe/H] plane \citep[e.g.,][]{1998A&A...338..161F, 2003A&A...410..527B, 2015ApJ...808..132H}. This high-$\alpha$ disk component is centrally concentrated and kinematically hot \citep[e.g.][]{2011ApJ...738..187L, 2012ApJ...753..148B, 2016ApJ...823...30B}, tracing the earlier phase of the Galactic disk formation \citep[e.g.][]{2013A&A...560A.109H, 2022Natur.603..599X}. While it largely represents the chemical counterpart of the geometrically defined thick disk \citep{1983MNRAS.202.1025G, 2008ApJ...673..864J}, the two are not structurally identical. The high-$\alpha$ disk exhibits a more complex configuration, with properties that vary strongly with both metallicity and stellar age \citep{2012ApJ...753..148B, 2013A&ARv..21...61R, 2025NatAs...9..101X}.  

Recent observations have revealed that the high-$\alpha$ disk stars follow a remarkably tight age--metallicity across the 8--13 Gyr age range \citep{2022Natur.603..599X}, suggesting efficient mixing and a highly regulated chemical-enrichment history. Nevertheless, the formation mechanism that sustains such a tight relation over such an extended timescale remains poorly understood. Unraveling this mechanism is crucial for gaining deeper insights into the early formation and evolution of the Galactic disk. 
An important clue may lie in whether the high-$\alpha$ disk is chemically smooth or contains identifiable substructures associated with distinct star-forming events. Resolving such structures requires examining the stellar population of the high-$\alpha$ disk in a high-dimensional chemical-abundance space.

Stellar chemical tagging has been proposed as a tool for distinguishing stellar families and identifying populations that share common birth origins or similar chemical-enrichment histories \citep{2002ARA&A..40..487F, 2004PASA...21..110B, 2015ApJ...807..104T, 2016ApJ...833..262H}. In its strongest form, chemical tagging aims to identify stars born in the same molecular cloud but subsequently dispersed throughout the Galaxy by dynamical evolution. The idea rests on the assumption that stars formed from the same cloud share similar chemical abundance patterns. This assumption is broadly supported by abundance measurements of star cluster members, which show star-to-star abundance dispersions of only about 0.02--0.03~dex for most elements up to Zn \citep[e.g.][]{2016ApJ...817...49B, 2026MNRAS.tmp.1588W}, although larger dispersions are often found for neutron-capture elements \citep{2026MNRAS.tmp.1588W}.



Chemical tagging of disk stars requires precise abundance measurements for various elements across large stellar samples \citep{2015ApJ...807..104T, 2016ApJ...816...10T}. Large-scale, high-resolution spectroscopic surveys such as APOGEE \citep{2017AJ....154...94M, 2017AJ....154..198Z, 2017arXiv171103234K} and GALAH \citep{2015MNRAS.449.2604D, 2025PASA...42...51B} have paved the way for such studies. Using APOGEE abundances derived with \textit{The Cannon} \citep{2015ApJ...808...16N}, \citet{2016ApJ...833..262H} demonstrated that members of stellar clusters and dwarf galaxies can be recovered as over-densities in high-dimensional chemical-abundance space using conventional clustering algorithms. Similar methods have also been used to identify candidate members of dissolved stellar clusters and chemically coherent groups across the Galactic disk \citep{2020MNRAS.496.5101P}. Studies based on GALAH abundances have reached similar conclusions \citep{2018MNRAS.473.4612K, 2020MNRAS.491.3374S}. In particular, recent studies have shown that neutron-capture elements can provide additional discriminating power in chemical-abundance space \citep{2024ApJ...972...69M, 2026MNRAS.tmp.1588W}.

On the other hand, although individual elemental lines are often blended at low spectral resolution, their collective and wavelength-dependent imprints remain encoded in the full spectrum. By leveraging physics-informed artificial intelligence (AI) models that effectively extract these weak yet informative signatures from the full spectrum, robust stellar abundances for over 20 elements, including neutron-capture species, can be derived from low-resolution spectra as well \citep{2017ApJ...849L...9T, 2019ApJS..245...34X, 2026ApJ..1002...22W}. This AI-driven breakthrough has enabled the derivation of individual elemental abundances for millions of stars from large-scale low-resolution spectroscopic surveys \citep{2019ApJS..245...34X, 2020ApJ...898...58W, 2024ApJS..273...19Z, 2025ApJS..279....5Z, 2026arXiv261004294V}. The reliability of low-resolution spectroscopic abundances has been corroborated through their application to characterizing chemically peculiar stars \citep{2020ApJ...898...28X, 2023ApJ...946..110Z} and through independent assessments of intrinsic abundance dispersion of the Galactic disk stars \citep{2026MNRAS.tmp.1588W}. However, whether the precision and dimensionality of low-resolution spectroscopic abundances are sufficient to resolve chemically coherent structures in the high-$\alpha$ disk remains largely unexplored.

In this work, we use elemental abundances derived from LAMOST low-resolution spectra to investigate the chemical structure of the high-$\alpha$ disk. To this end, we construct a sample of 87,275 high-$\alpha$ disk red giant stars with high-quality abundance measurements, and apply the Density-Based Spatial Clustering of Applications with Noise algorithm \citep[DBSCAN;][]{ester1996density} to identify chemically coherent groups in an 18-dimensional abundance space. The effectiveness of our approach is assessed through comparative analyses using abundances from the high‑resolution APOGEE survey as well as from mock catalogs. Our results not only offer new insights into the formation history of the high‑$\alpha$ disk, but also provide a critical benchmark for evaluating both the potential and the limitations of low‑resolution stellar abundances in chemical tagging studies.


This paper is organized as follows. Section~\ref{sec:label} describes the data and sample selection. Section~\ref{sec:method} introduces the chemical-distance metric, DBSCAN methodology, and validation tests. Section~\ref{sec:result} presents the clustering results, the identification and characterization of chemical families. Section~\ref{sec:discussion} discusses the implications and limitations of our findings. Finally, Section~\ref{sec:conclusion} summarizes our main conclusions.

\section{data} 
\label{sec:label}
We use the LAMOST DR9 stellar abundance catalog of \citet{2025ApJS..279....5Z}, which provides homogeneous abundance measurements of 22 elements — C, N, O, Na, Mg, Al, Si, Ca, Ti, Cr, Mn, Fe, Ni, Sr, Y, Zr, Ba, La, Ce, Nd, Sm, and Eu — for about 3.6 million FGK stars. The abundances are derived from LAMOST low-resolution spectra with the DD-Payne, a physical-prior imposed data-driven neural network spectral modeling approach \citep{2019ApJS..245...34X}. For spectra with ${\rm S/N}>50$, typical internal uncertainties are $\simeq$0.05~dex for most elemental abundances with atomic number no larger than Ni, while the uncertainties for heavier, neutron-capture elements are 0.1-0.2 dex. In particular, given the same ${\rm S/N}$, red giant stars yield more precise abundances than dwarfs, owing to their stronger spectral features. 

For the current purpose, we construct a high-quality sample of red giant stars by applying the following selection criteria:
\[ \begin{cases} &S/N_g > 40,
\\ &T_{\rm eff}<5600~\text{K},
\\ & 2<\log g <3.5, 
\\ &\text{[Fe/H]}>-1, 
\end{cases} \]
where $S/N_g$ is the spectral signal-to-noise ratio in the SDSS $g$ band. These cuts yield a sample of 355,909 red giant stars.

\begin{figure}[htbp]
		\centering
    \includegraphics[width=1\linewidth]{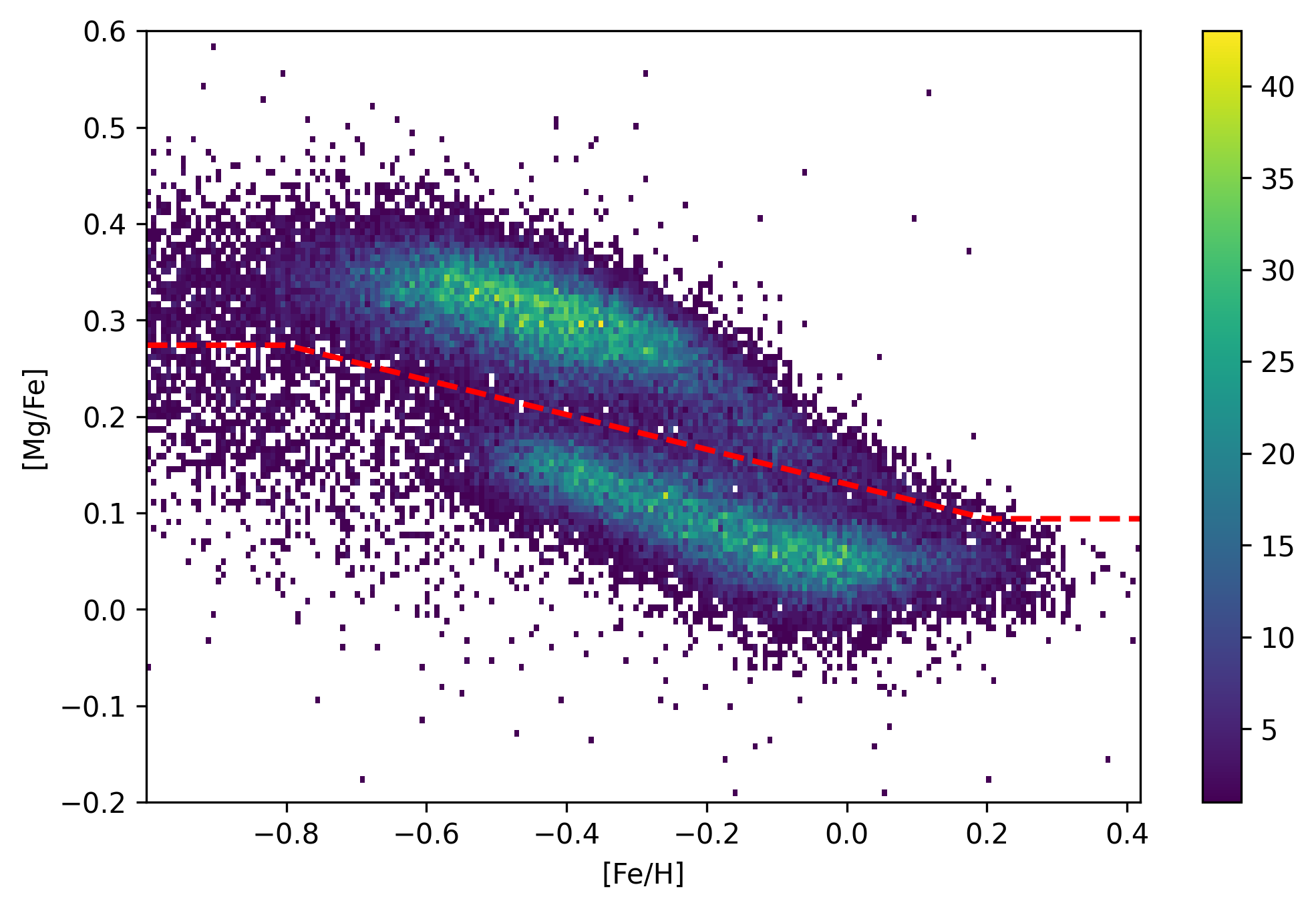}
		\caption{Density distribution plot of the high-quality red giant star sample in the $[\mathrm{Mg/Fe}]$--$[\mathrm{Fe/H}]$ plane. The red dashed line marks the adopted boundary to separate the high-$\alpha$ disk population from the low-$\alpha$ population. Stars above this boundary are selected as the high-$\alpha$ disk sample.}
       \label{fig:thickdisksample}
\end{figure}
From this high-quality red giant sample, we further select high-$\alpha$ disk stars according to the $\mathrm{[Mg/Fe]}$–$\mathrm{[Fe/H]}$ plane. We adopt an empirical boundary that traces the low-density region between the high-$\alpha$ and low-$\alpha$ sequences, as shown in Figure~\ref{fig:thickdisksample}. 
Specifically, the boundary is defined as,
\begin{equation}
\mathrm{[Mg/Fe]}=
\begin{cases}
0.274, & \mathrm{[Fe/H]}<-0.8,\\
-0.18\,\mathrm{[Fe/H]}+0.13, & -0.8\leq\mathrm{[Fe/H]}\leq0.2,\\
0.094, & \mathrm{[Fe/H]}>0.2.
\end{cases}
\end{equation}
Stars lying above this boundary are classified as high-$\alpha$ disk stars. A total of 87,275 stars lying above this boundary are selected as the high-$\alpha$ disk sample and are subsequently used to construct the multi-dimensional chemical abundance space for our DBSCAN analysis.

The chemical-abundance space is constructed using $\mathrm{[Fe/H]}$ and the abundance ratios of C, N, Mg, Al, Si, Ca, Ti, Cr, Mn, Ni, and neutron-capture elements (Y, Zr, Ba, La, Ce, Nd, Sm, and Eu). Here we exclude O, Na, and Sr because of their large measurement uncertainties owing to either poor spectra features or poor training labels of the DD-Payne \citep{2025ApJS..279....5Z}. In addition, dredge-up processes during stellar evolution may alter the surface abundances of C and N. To mitigate the impact of this effect, we combine C and N into a single chemical dimension, $\mathrm{[(C+N)/Fe]}$, according to the following equation:
\begin{equation}
[\rm C+N/Fe]
=
\log_{10}
\left(
\frac{n_{\rm C}+n_{\rm N}}{n_{\rm H}}
\right)
-
\log_{10}
\left(
\frac{n_{\rm C_{\odot}}+n_{\rm N_{\odot}}}{n_{\rm H_{\odot}}}
\right)-[Fe/H],
\end{equation}

\begin{equation}
\frac{n_{\rm C}+n_{\rm N}}{n_{\rm H}}
=
10^{[\rm C/H]}
\left(\frac{n_{\rm C}}{n_{\rm H}}\right)_{\odot}
+
10^{[\rm N/H]}
\left(\frac{n_{\rm N}}{n_{\rm H}}\right)_{\odot},
\end{equation}
where the solar reference abundances are adopted from \citet{2009ARA&A..47..481A}.
Accordingly, the uncertainty in $[\mathrm{(C+N)/Fe}]$ is estimated via error propagation of the C and N abundance determinations.


Combining the DD-Payne stellar parameters, radial velocities are provided in the LAMOST official release \footnote{http://www.lamost.org/dr9/}. With photometry and astrometry information from Gaia DR3 \citep{2021A&A...649A...1G}, we also derive the stellar distance and orbital parameters, following the same approach as introduced in \citet{2022Natur.603..599X}. The orbital parameters are derived with {\sc Galpy} \citep{2015ApJS..216...29B}, using the default Milky Way potential 'MWPotential2014'. 

\section{Method}
\label{sec:method}
To identify chemically coherent stellar structures in the Galactic thick disk, we perform clustering analysis directly in high-dimensional chemical abundance space. 
In this framework, the chemical similarity between stars is quantified using an uncertainty-weighted distance metric, and chemically coherent groups are subsequently identified with the DBSCAN \citep{ester1996density} method.

\subsection{Uncertainty-weighted Distance Metric}
In the chemical abundance space, each star forms a vector constructed from its elemental abundance ratios
\begin{equation}
\boldsymbol{X} = ([\mathrm{Fe/H}], [\mathrm{(C+N)/Fe}], [\mathrm{Mg/Fe}], [\mathrm{Al/Fe}] \ldots).
\end{equation}

To ensure that different abundance dimensions contribute comparably to the distance metric, preventing elements with intrinsically larger numerical dispersions from dominating the clustering results, we apply standard normalization to each abundance dimension. 
For each element $k$, the normalized abundance is defined as
\begin{equation}
X'_{i,k} = \frac{X_{i,k} - \mu_k}{s_k},
\end{equation}
where $X_{i,k}$ is the measured abundance of the $k$-th element for star $i$, while $\mu_k$ and $s_k$ denote the mean and standard deviation of that abundance dimension over the full sample, respectively.
The corresponding abundance uncertainties are also normalized consistently as
\begin{equation}
\sigma'_{i,k} = \frac{\sigma_{i,k}}{s_k},
\end{equation}
where $\sigma_{i,k}$ denotes the abundance uncertainty of the $k$-th element for star $i$.

Unlike the standard Euclidean distance commonly adopted in DBSCAN, we define a customized distance metric that incorporates measurement uncertainties in the chemical abundances. 
Elements with larger measurement uncertainties are assigned lower weights, ensuring that the distance metric is dominated by well-constrained abundance information rather than noise. 
The pairwise chemical distance between stars $i$ and $j$ is defined as
\begin{equation}
D_{ij} = \sqrt{
\frac{
\sum_{k=1}^{N_{dim}} w_{ij,k} \left( X_{i,k} - X_{j,k} \right)^2
}{
\sum_{k=1}^{N_{dim}} w_{ij,k}
}
},
\end{equation}
where $X_{i,k}$ denotes the normalized abundance of the $k$-th element for star $i$, and $N_{dim}$ is the total number of chemical dimensions.  
The weight $w_{ij,k}$ is defined from the abundance uncertainties as
\begin{equation}
w_{ij,k} = \frac{1}{{\sigma'}_{i,k}^2 + {\sigma'}_{j,k}^2},
\end{equation}
where ${\sigma'}_{i,k}$ and ${\sigma'}_{j,k}$ denote the normalized uncertainties of the $k$-th element for stars $i$ and $j$, respectively.

\subsection{DBSCAN Clustering Method}
Based on the pairwise chemical distances defined above, we apply the DBSCAN clustering algorithm to identify chemically coherent structures in high-dimensional abundance space.
DBSCAN is a density-based clustering method that groups points according to the local density of the data.
It relies on two parameters: a neighborhood radius ($\epsilon$) and a minimum number of samples ($\mathrm{MinPts}$). Points with at least $\mathrm{MinPts}$ neighbors within $\epsilon$ are defined as core points, which form the basis of clusters. Clusters are then constructed by connecting points that are within the $\epsilon$-neighborhood of these core points, which we interpret as candidate 'chemical groups'. Stars that are not assigned to any cluster are labeled as noise under the adopted DBSCAN parameters.

Compared to partition-based methods (e.g., $k$-means), DBSCAN does not require a predefined number of clusters and is therefore well suited for exploring unknown substructures in the chemical abundance space. This makes it particularly suitable for our purpose.

\subsection{Comparison of DBSCAN Clustering Results between APOGEE and LAMOST}

To assess the sensitivity of our clustering method to spectroscopic abundance measurements and the choice of chemical dimensions, we perform a comparative analysis between APOGEE DR17 and LAMOST DR9. We construct a high-quality stellar sample by cross-matching the APOGEE DR17 catalog \citep{2022ApJS..259...35A} with the LAMOST DR9 DD-Payne measurements \citep{2025ApJS..279....5Z}. We select unique stars satisfying the following criteria for the LAMOST observation:
\[ \begin{cases} &S/N_g > 40,
\\ &3800~\text{K}<T_{\rm eff}<5600~\text{K},
\\ & \log g <3.8, 
\\ &\text{[Fe/H]}>-1.
\end{cases} \]
These criteria yield a sample of 44,657 stars.

To enable a direct comparison between APOGEE and LAMOST, we first define a common chemical-abundance space consisting of elements measured in both surveys, including $[\mathrm{Fe/H}]$, $[\mathrm{C+N/Fe}]$, $[\mathrm{Mg/Fe}]$, $[\mathrm{Al/Fe}]$, $[\mathrm{Si/Fe}]$, $[\mathrm{Ca/Fe}]$, $[\mathrm{Ti/Fe}]$, $[\mathrm{Cr/Fe}]$, $[\mathrm{Mn/Fe}]$, and $[\mathrm{Ni/Fe}]$. 
For the subsequent analysis of the LAMOST sample, we further extend the chemical-abundance space by incorporating additional neutron-capture elements, including Y, Zr, Ba, La, Ce, Nd, Sm, and Eu, which are not fully available in APOGEE. 
This comparison examines how including these additional abundance dimensions affects the chemical-distance distribution.

\begin{figure}[htbp]
	\centering
    \includegraphics[width=1\linewidth]{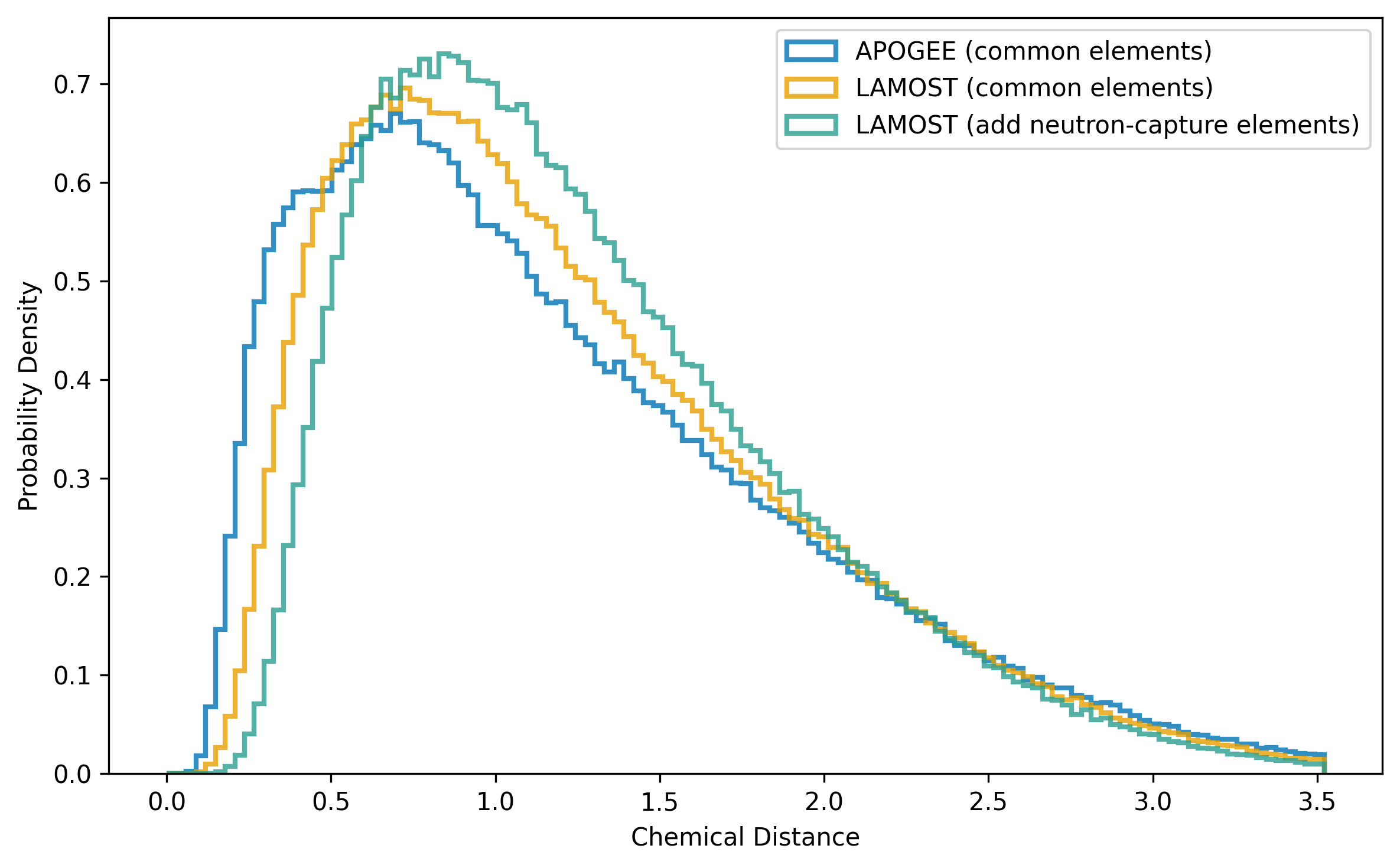}
		\caption{Distribution of pairwise distances derived using abundances of the APOGEE (blue), LAMOST (orange), LAMOST-add neutron-capture elements (green) for the LAMOST-APOGEE common star sample.}
       \label{fig:lam_apo_distance}
\end{figure}

Figure~\ref{fig:lam_apo_distance} shows the distributions of pairwise chemical distances for APOGEE and LAMOST. The APOGEE sample exhibits a more compact chemical-distance distribution, consistent with the generally higher precision of its abundance measurements. In contrast, the LAMOST sample distribution is broader and shifted toward larger distances, likely due to larger measurement errors in the derived abundances.
When additional neutron-capture elements are included, the distance distribution for LAMOST shifts further toward larger values and becomes slightly broader. This behavior is expected, as the neutron capture element abundances exhibit larger variances compared to other elements.

\begin{figure*}[htbp]
		\centering
        \includegraphics[width=1\linewidth]{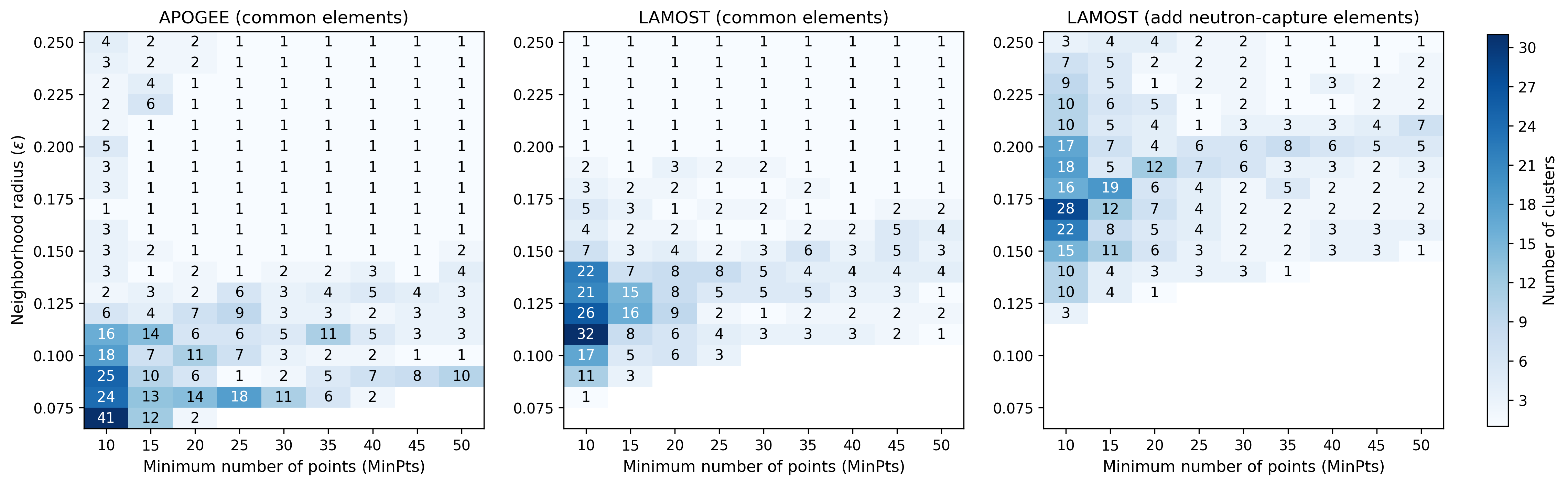}
		\caption{Number of non-noise clusters identified by \text{DBSCAN} for different parameter combinations. The three panels correspond to the APOGEE sample (left), the LAMOST sample using the common element set (middle), and the LAMOST sample including additional neutron-capture elements (right). The horizontal axis shows $\mathrm{MinPts}$, and the vertical axis shows the neighborhood radius $\epsilon$. The color scale and the annotated numbers indicate the resulting number of clusters, $N_{\rm clusters}$, for each parameter combination. }
       \label{fig:clusternumbers}
\end{figure*}

\begin{figure*}[htbp]
		\centering
        \includegraphics[width=1\linewidth]{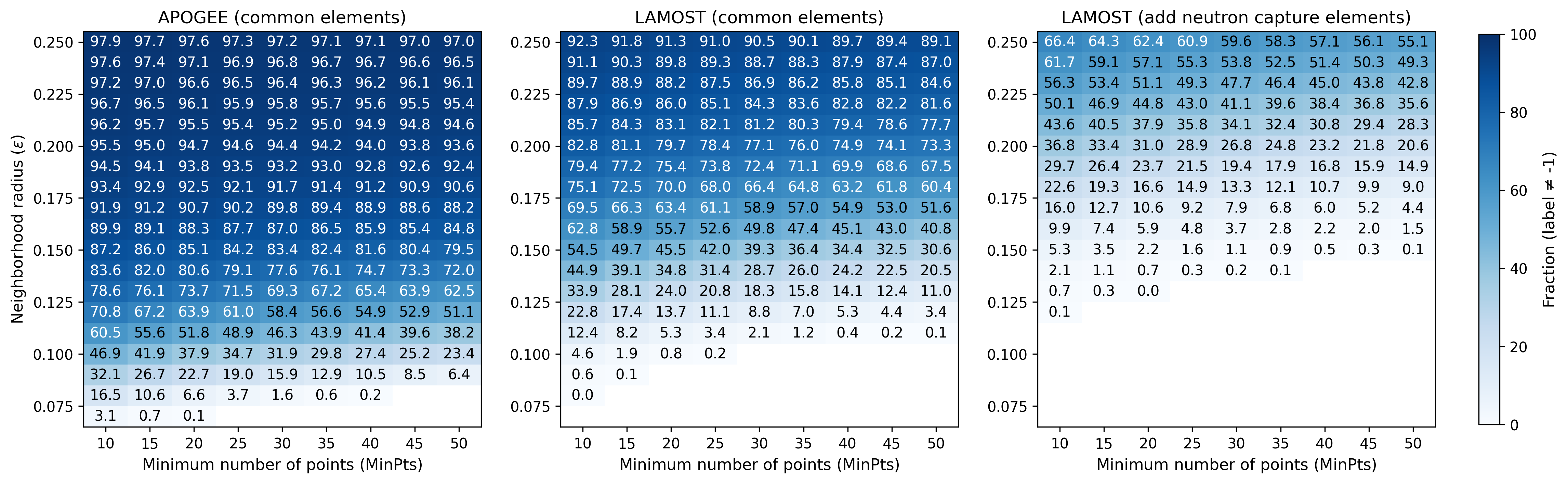}
		\caption{Same as Figure~\ref{fig:clusternumbers}, but showing the fraction of stars assigned to non-noise clusters by \text{DBSCAN}. The color scale and annotated values indicate the fraction of clustered stars for each parameter combination.}
       \label{fig:fraction}
\end{figure*}

We then apply DBSCAN to the three abundance datasets. We explore $\epsilon$ from 0.07 to 0.25 in steps of 0.01 and $\mathrm{MinPts}$ from 10 to 50 in steps of 5. 
Figure~\ref{fig:clusternumbers} shows the number of non-noise clusters for each parameter combination, and Figure~\ref{fig:fraction} shows the proportion of the clustered stars. 
These figures show that APOGEE abundances can effectively identify stellar clusters at $\epsilon<0.125$, whereas LAMOST abundances require larger values of $\epsilon$ ($0.09\lesssim\epsilon\lesssim0.16$). When neutron-capture elements are included, the corresponding range shifts to even larger values ($0.125\lesssim\epsilon\lesssim0.25$). These trends are expected, as the pairwise chemical distances are strongly affected by measurement uncertainties. 

Nevertheless, Figure~\ref{fig:fraction} shows that the fractions of clustered stars remain comparable across the three abundance sets. Specifically, APOGEE abundances yield clustered fractions of approximately 60--70\% at $\epsilon<0.125$, while LAMOST abundances yield similar fractions at $0.09\lesssim\epsilon\lesssim0.16$. Adding neutron-capture elements yields a clustered fraction of around 60\% at $0.125\lesssim\epsilon\lesssim0.25$. Notably, both APOGEE and LAMOST abundances yield clustered fractions exceeding 90\% at $\epsilon\simeq0.25$. This is an artifact of an excessively large $\epsilon$, which causes otherwise distinguishable populations to merge into a single dominant cluster rather than being identified as separate chemical structures. These results suggest that, despite their larger measurement uncertainties, LAMOST abundances can yield clustering results comparable to those obtained from APOGEE abundances in terms of both the number of identified clusters and the fraction of stars assigned to clusters.


\subsection{Test on mock data}
To test whether DBSCAN can artificially divide an intrinsically chemistry-homogeneous stellar population into multiple clusters, we construct a homogeneous mock sample based on the observed high-$\alpha$ disk stars. The mock sample contains the same number of stars as the observed sample and retains their individual abundance uncertainties.

For each element $k$, we assign all mock stars an identical intrinsic abundance, defined as the median abundance of the observed sample. We then generate mock abundances by adding independent Gaussian perturbations according to the measured abundance uncertainties of individual stars:
\begin{equation}
X_{i,k}^{\rm mock}
=
\widetilde{X}_k
+
\sigma_{i,k}z_{i,k},
\qquad
z_{i,k}\sim\mathcal{N}(0,1),
\end{equation}
where $\widetilde{X}_k$ is the median abundance of the observed high-$\alpha$ disk sample for element $k$, $\sigma_{i,k}$ is the measured abundance uncertainty of star $i$, and $z_{i,k}$ is an independent standard Gaussian random variable. The mock abundances and their uncertainties are subsequently standardized using the means and standard deviations of the real high-$\alpha$ disk sample.

\begin{figure*}[htbp]
		\centering
        \includegraphics[width=1\linewidth]{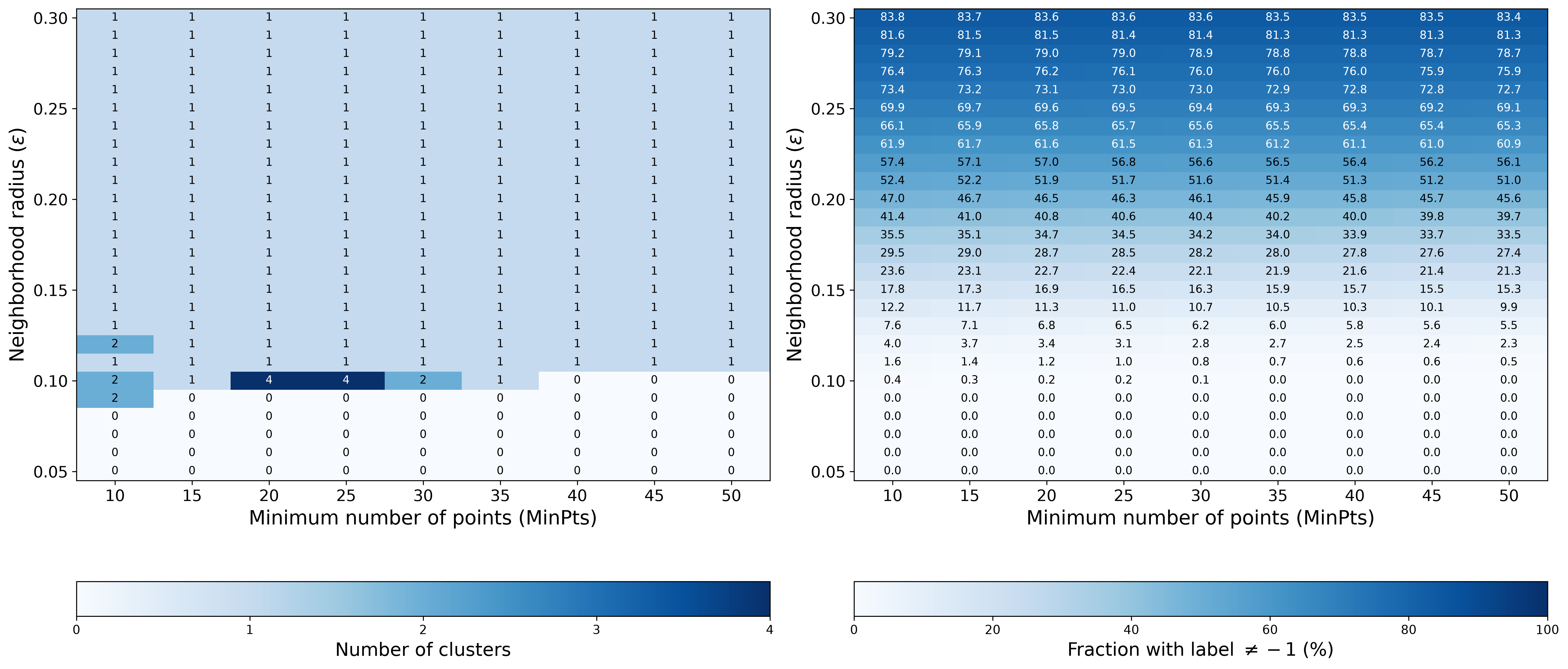}
		\caption{DBSCAN clustering results for the chemically homogeneous mock sample. The left panel shows the number of clusters identified for different parameter combinations, while the right panel shows the corresponding fraction of clustered stars. }
       \label{fig:mock_results}
\end{figure*}

We apply the same uncertainty-weighted distance metric and DBSCAN clustering procedure to the mock sample using clustering parameters with $\epsilon$ ranging from 0.05 to 0.30 in steps of 0.01 and $\mathrm{MinPts}$ ranging from 10 to 55 in steps of 5. Figure~\ref{fig:mock_results} shows the number of detected clusters and the fraction of stars assigned to clusters for different parameter combinations. For most parameter combinations that yield non-noise clusters, DBSCAN identifies a single cluster. Multiple clusters, with a maximum of four, appear only within a narrow range of small $\epsilon$, whereas more restrictive parameter combinations yield no clusters. 

The clustering results suggest that DBSCAN does not systematically fragment an intrinsically chemically homogeneous stellar population into numerous spurious clusters. The major difference between the observed and mock sample is the astrophysical correlation among different elements: in the observed sample, elemental abundances exhibit astrophysical correlations arising from shared nucleosynthetic origins, leaving coherent imprints on their abundance patterns. In contrast, these astrophysical correlations are absent in the mock sample, where all stars are assigned identical intrinsic abundances for each element, followed by independent Gaussian perturbations based on their measurement uncertainties. Our test therefore suggests that the coherent multi-element abundance patterns present in the observed sample contain important information for chemical tagging. This is consistent with the findings of \citet{2022ApJ...927..209T}, who highlighted the importance of astrophysical correlations among elemental abundances for chemical tagging. Together, these findings underscore the potential value of coherent multi-element abundance patterns for chemical tagging and for identifying chemical structures in high-dimensional abundance space.


\section{Results}
\label{sec:result}
\subsection{DBSCAN Clustering in High-Dimensional Chemical Space}

\begin{figure*}[htbp]
		\centering
        \includegraphics[width=1\linewidth]{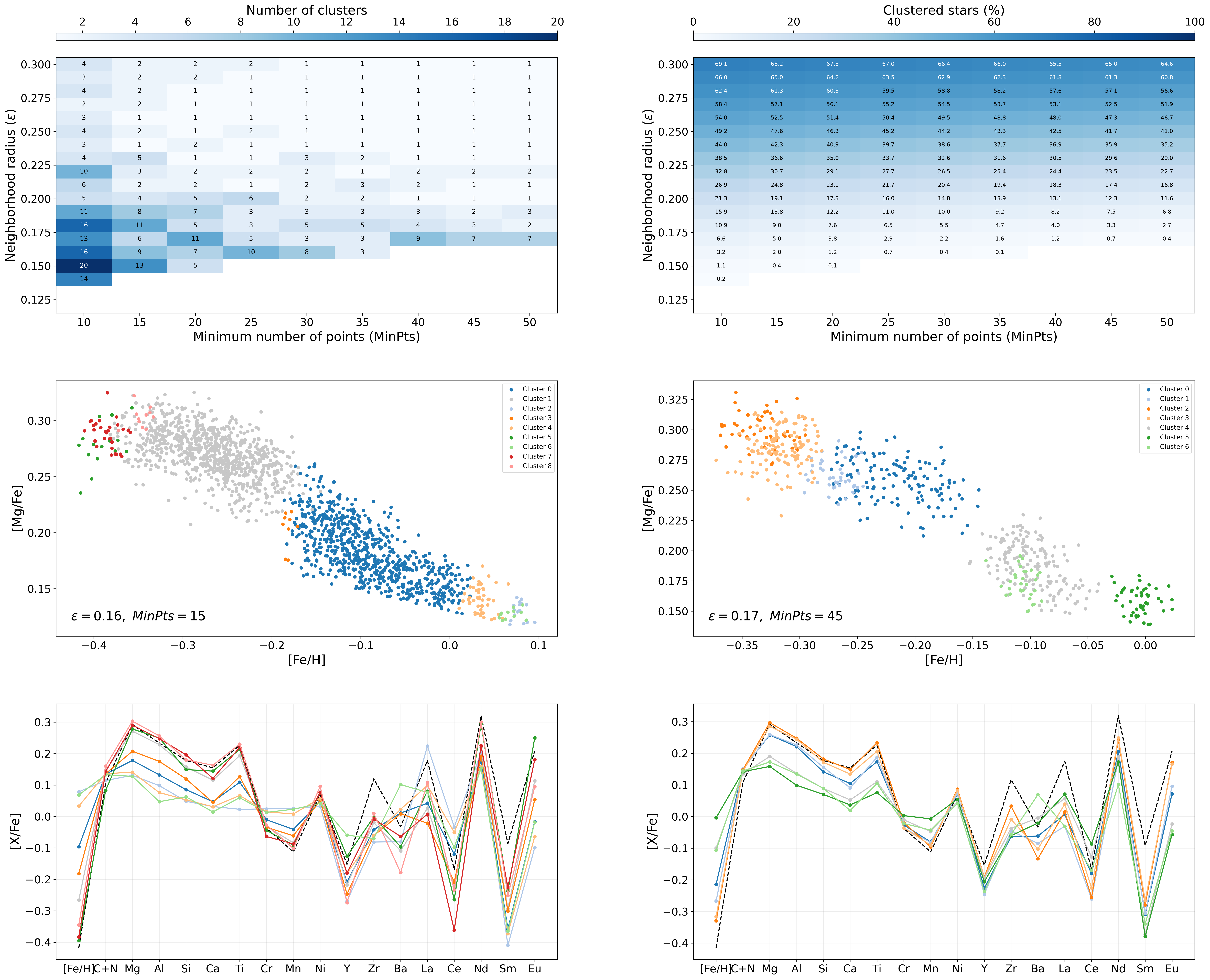}
		\caption{DBSCAN clustering results with different parameter combinations. \textit{Top panels:} Number of clusters (left) and fraction of clustered stars (right) as a function of the DBSCAN parameters $\epsilon$ and $\mathrm{MinPts}$. 
        \textit{Middle panels:} Clustering results projected onto the $[\mathrm{Mg/Fe}]$--$[\mathrm{Fe/H}]$ plane for two representative parameter choices: $\epsilon=0.16$, $\mathrm{MinPts}=15$ (left), and $\epsilon=0.17$, $\mathrm{MinPts}=45$ (right).
        \textit{Bottom panels:} Median chemical abundance patterns of the identified clusters. The dashed black curve represents the median abundance pattern of the unclustered stars. }
       \label{fig:sample}
\end{figure*}

As the clustering results produced by DBSCAN depend on the adopted parameter choices, we first systematically investigate how the identified chemical structures vary across different combinations of $\epsilon$ and $\mathrm{MinPts}$.
After extensive experiments, we adopted an appropriate parameter range, with $\epsilon$ varying from 0.12 to 0.3 in steps of 0.01, and $\mathrm{MinPts}$ ranging from 10 to 50 in steps of 5. 

Figure~\ref{fig:sample} shows the number of identified clusters (top-left panel) and the fraction of stars assigned to non-noise clusters (top-right panel) for different parameter combinations. 
The results indicate that the high-$\alpha$ thick-disk sample exhibits significant chemical substructures in high-dimensional abundance space. 
At a relatively small $\epsilon$, DBSCAN preferentially identifies only the most compact and chemically distinct overdensities, resulting in a relatively large number of small clusters but a low clustered fraction.
As $\epsilon$ increases, more stars become connected in chemical-abundance space, leading to a rapid increase in the clustered fraction while progressively merging previously separated structures.
This behavior indicates that the chemical structures in the thick disk are not completely isolated from each other, but instead exhibit a certain degree of continuity and hierarchical organization in abundance space.

The parameter $\mathrm{MinPts}$ mainly controls the minimum local density required for a structure to be identified as a cluster.
Larger $\mathrm{MinPts}$ suppress sparse overdensities and generally reduce both the number of identified clusters and the clustered fraction, suggesting that many chemically coherent structures in the thick disk contain only a relatively small number of stars in our sample. 
Although the exact clustering properties vary with the adopted DBSCAN parameters, chemically coherent overdensities can be consistently identified over a broad parameter range, indicating a robust clustering signal beyond random noise.

Across the explored parameter combinations, a total of 447 non-noise DBSCAN clusters are identified. 26,951 stars were classified as noise across all parameter combinations. 
The middle and bottom panels of Figure~\ref{fig:sample} present two representative examples of the chemical clusters identified by DBSCAN. The middle panels show the distribution of stars for different clusters in the $[\mathrm{Mg/Fe}]$--$[\mathrm{Fe/H}]$ plane. Some clusters overlap substantially in this conventional two-dimensional abundance space, suggesting that clustering in the high-dimensional chemical abundance space provides superior separation. The bottom panels display the median abundance patterns of different clusters. The dashed black curve shows the overall median abundance pattern of the unclustered population. Different clusters exhibit clear differences for most of the elemental dimensions.

\subsection{Identification of Chemical Families through Different Parameter Grids}
\label{sec:Identification of Chemical Families}
Since DBSCAN is sensitive to the choice of $\epsilon$ and $\mathrm{MinPts}$, different parameter combinations may identify clusters with similar chemical abundance patterns. We therefore compare clusters identified across different parameter combinations and merge chemically similar clusters into chemical families, while aiming to preserve distinct abundance patterns.

\begin{figure*}[htbp]
	\centering
        \includegraphics[width=1\linewidth]{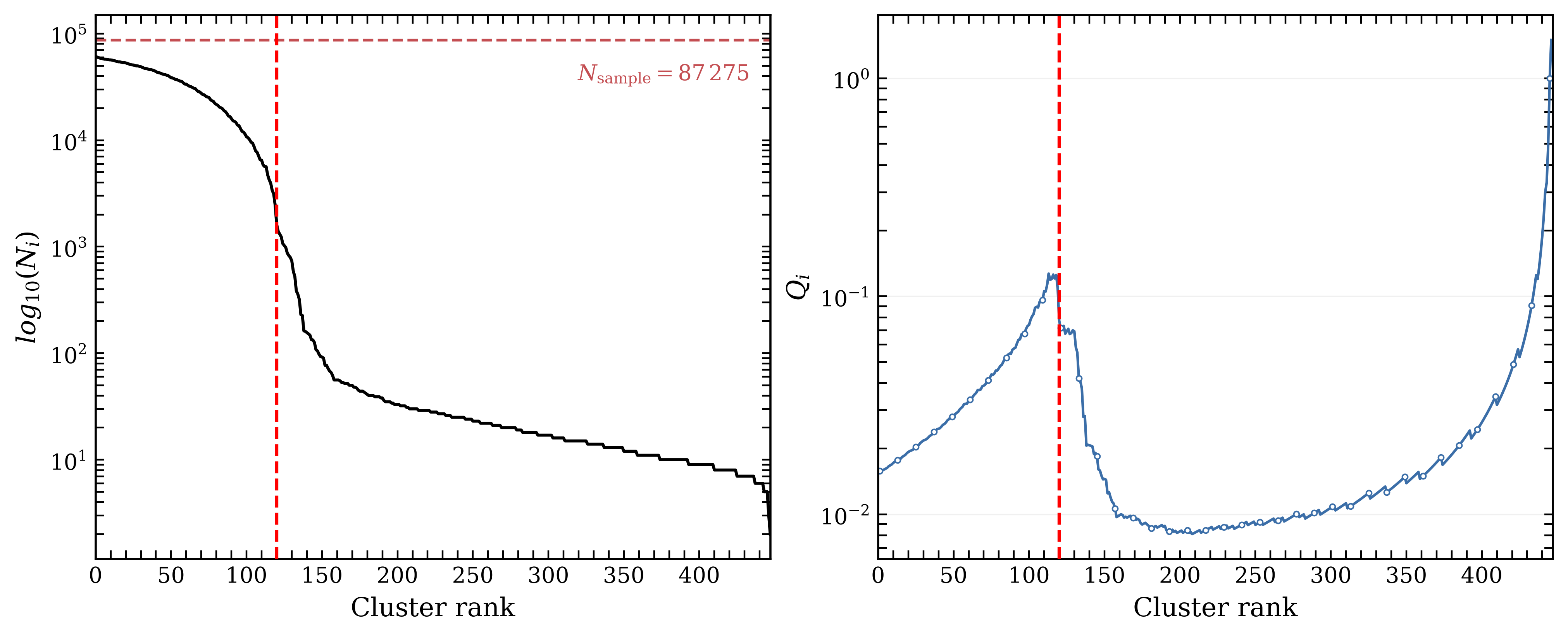}
	\caption{Cluster-size distribution and diagnostic for identifying potentially percolating structures. Left: $\log_{10}(N_i)$ as a function of cluster rank, where $N_i$ is the number of member stars in the $i$-th cluster, with clusters sorted in descending order of membership. The horizontal red dashed line indicates the total number of stars in the high-$\alpha$ sample, $N_{\rm sample}=87\,275$, and the vertical red dashed line marks rank 120. Right: Ratio $Q_i=N_i/\sum_{j>i}N_j$, where the denominator is the summed memberships of all lower-ranked clusters. The ratio measures the size of an individual cluster relative to the combined memberships of smaller clusters. The vertical red dashed line indicates the same screening boundary. The transition in the cluster-size distribution motivates the exclusion of the 120 largest clusters before cross-grid merging.}
       \label{fig:percolation}
\end{figure*}

Prior to cross-grid merging, we apply a screening procedure to mitigate potential percolation effects in the DBSCAN results. As $\epsilon$ increases, neighboring dense regions may become connected through intermediate stars, producing large clusters that span broad regions of chemical-abundance space and potentially obscure smaller-scale chemical substructures. To identify potentially percolating clusters, we examine the cluster-size distribution across all DBSCAN parameter combinations (Figure~\ref{fig:percolation}). In the left panel, clusters are ranked in descending order of their membership, with $N_i$ denoting the number of stars in the $i$-th ranked cluster. In the right panel, we calculate the ratio
\begin{equation}
Q_i = \frac{N_i}{\sum_{j>i} N_j},
\end{equation}
where the denominator represents the total number of cluster memberships summed over all clusters ranked below the $i$-th cluster. This ratio measures the size of an individual cluster relative to the combined memberships of all smaller clusters. A large value indicates that a single cluster contributes substantially to the remaining lower-ranked cluster population, whereas a small value indicates that its membership is relatively minor compared with that population.

As shown in Figure~\ref{fig:percolation}, the cluster-size distribution exhibits a pronounced transition around rank 120, accompanied by a local peak in $Q_i$ followed by a sharp decline. Based on these features, we adopt rank 120 as an empirical screening boundary and exclude the 120 largest clusters before cross-grid merging. We also tested a more restrictive boundary at rank 150, which yielded a comparable number of chemical families, but substantially reduced the fraction of unique stars included in the resulting catalog. We therefore adopt rank 120 as the fiducial boundary, motivated by the observed transition in the cluster-size distribution and to avoid the substantial loss of stellar coverage associated with a more restrictive cut.

Following this screening procedure, all retained DBSCAN clusters are collected into a unified catalog for cross-grid comparison. To quantify their chemical similarity, we characterize each cluster, $C$, by its median abundance vector,
\begin{equation}
\tilde{\mathbf{X}}_C =
(\tilde{X}_{C,1},\tilde{X}_{C,2},\ldots,\tilde{X}_{C,18}),
\end{equation}
where $\tilde{X}_{C,k}$ denotes the median abundance of the $k$-th chemical dimension among the member stars of cluster $C$. The uncertainty of the median abundance is estimated from the median absolute deviation (MAD) as
\begin{equation}
\sigma_{C,k}=1.253\times1.4826
\frac{{\rm MAD}_{C,k}}{\sqrt{N_C}}.
\end{equation}
Here, $N_C$ is the number of stars in the cluster, and MAD is the median absolute deviation of the stellar abundances from the cluster median. The factor 1.4826 comes from the relation $\mathrm{MAD}\approx0.6745\sigma$ for a Gaussian distribution and converts MAD into an estimate of the standard deviation. The factor 1.253 is approximately equal to $\sqrt{\pi/2}$ and accounts for the larger sampling uncertainty of the median compared to the mean for independent Gaussian data.

Using these median abundance vectors and their uncertainties, we define the chemical distance between two clusters, $i$ and $j$, as
\begin{equation}
D(i,j)=
\sqrt{
\sum_{k=1}^{N_{\rm dim}}
\frac{(\tilde{X}_{i,k}-\tilde{X}_{j,k})^2}
{
\sigma_{i,k}^{2}+\sigma_{j,k}^{2}}
},
\end{equation}

We then merge chemically similar clusters using a complete-linkage procedure to combine clusters identified under different DBSCAN parameter combinations into families. For complete-linkage, the maximum distance between any two clusters within a family must be smaller than the merging threshold $D_{\rm crit}$. 
\begin{equation}
\max D_{ij} < D_{\rm crit}.
\end{equation}

\begin{figure}[htbp]
	\centering
        \includegraphics[width=1\linewidth]{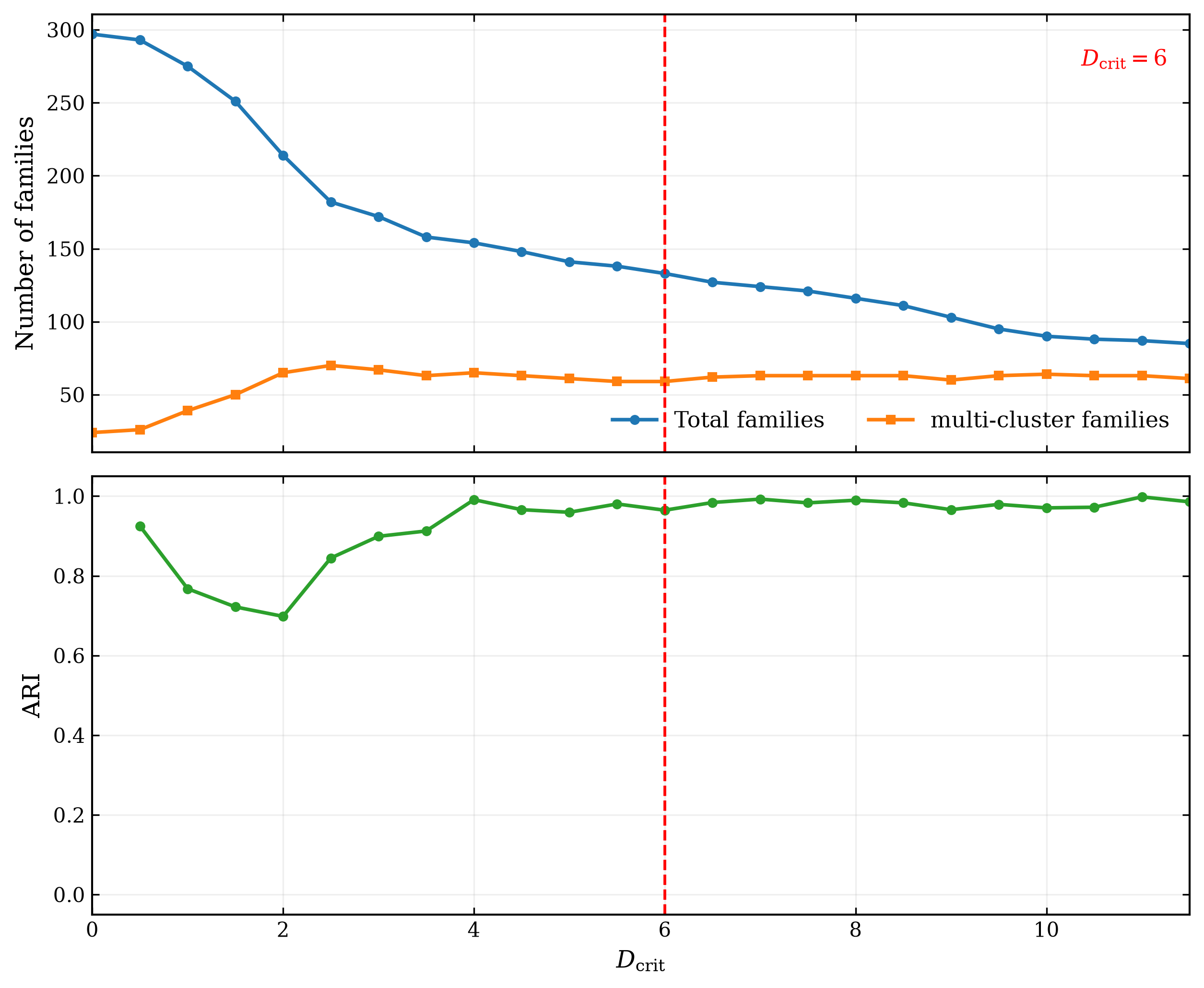}
	\caption{Dependence of chemical-family identification on the merging threshold $D_{\rm crit}$. The vertical red dashed line marks the adopted threshold, $D_{\rm crit}=6$. Top: Total number of chemical families (blue circles) and the number of chemical families composed of multiple DBSCAN clusters (orange squares) as functions of $D_{\rm crit}$. Bottom: Adjusted Rand Index (ARI) between family partitions obtained at adjacent threshold values, quantifying the stability of the family assignments as the merging threshold changes. }
       \label{fig:dcrit}
\end{figure}

To select $D_{\mathrm{crit}}$, we vary it from 0 to 12 in steps of 0.5. The top panel of Figure~\ref{fig:dcrit} reports the total number of families and the number of families composed of multiple DBSCAN clusters. The number of multi-cluster families initially increases with $D_{\mathrm{crit}}$, but subsequently fluctuates around 60--65 for $D_{\mathrm{crit}}>3$, indicating a relatively stable number of multi-cluster families over this threshold range. We further quantify the stability of the family partition using the adjusted rand index (ARI) between partitions of the retained DBSCAN clusters obtained at adjacent threshold values  \citep{hubert1985comparing}. An ARI of 1 indicates identical partitions, whereas an ARI of 0 indicates agreement comparable to that expected by chance. Higher ARI values ($\sim 1$) therefore indicate greater stability of the family partitions against changes in the merging threshold. The bottom panel of Figure~\ref{fig:dcrit} shows that the ARI increases substantially from the low-threshold regime and remains generally above 0.9 for $D_{\rm crit}\geq 4$. We adopt $D_{\rm crit}=6$ as our fiducial merging threshold. This value lies within the regime of high partition stability. Compared with a more restrictive threshold such as $D_{\rm crit}=4$, adopting $D_{\rm crit}=6$ does not substantially change the total number of chemical families, while allowing additional clusters with similar chemical abundance patterns to be merged into the same family. This choice therefore balances the stability of the resulting partition with our aim of consolidating clusters with similar chemical abundance patterns.

Using the adopted merging threshold of $D_{\rm crit}=6$, we identify a total of 133 chemical families. Among them, 59 families are formed by merging two or more chemically similar DBSCAN clusters identified across different parameter combinations, while the remaining 74 families consist of a single DBSCAN cluster. The 133 chemical families collectively contain 7298 stellar memberships, corresponding to 4642 unique stars, approximately 5.3\% of the high-$\alpha$ disk sample. Among these stars, 3,087 (66.3\%) are associated with a single chemical family. Note that although the identified families are characterized by different elemental abundance patterns, their stellar memberships are not necessarily mutually exclusive due to the abundance errors of individual stars.


We compile a catalog of all 133 chemical families, providing their spatial, kinematic, and chemical properties. For each family, the catalog includes the number of member stars, the number of constituent DBSCAN clusters. The spatial and orbital properties are characterized by the median Galactocentric cylindrical radius ($R$), vertical position ($Z$), angular momentum ($L_Z$), and vertical action ($J_Z$). The chemical properties are represented by the median abundances of all 18 chemical dimensions, together with the uncertainties of the medians and the robust abundance dispersions. Table~\ref{tab:chemical_families} presents the ten largest chemical families, ranked by the number of member stars, while the complete catalog is provided in machine-readable form.

\begin{deluxetable*}{cccccccccccc}
\tablecaption{Spatial, Kinematic, and Chemical Properties of the Ten Largest Chemical Families\label{tab:chemical_families}}
\tabletypesize{\scriptsize}
\tablewidth{0pt}
\tablehead{
\colhead{Family ID} &
\colhead{$N_{\rm stars}$} &
\colhead{$N_{\rm clusters}$} &
\colhead{$R$} &
\colhead{$Z$} &
\colhead{$L_Z$} &
\colhead{$J_Z$} &
\colhead{[Fe/H]} &
\colhead{$\cdots$}&
\colhead{[Mg/Fe]} &
\colhead{[Al/Fe]} &
\colhead{$\cdots$}
}
\startdata
0 & 1646 & 5  & 8.102 & 0.649 & 1435.604 & 30.418
& $-0.272\pm0.002\;(0.055)$
& $\cdots$
& $0.271\pm0.001\;(0.023)$
& $0.230\pm0.001\;(0.035)$
& $\cdots$ \\
1 & 1378 & 9  & 8.139 & 0.473 & 1537.301 & 17.360
& $-0.095\pm0.002\;(0.058)$
& $\cdots$
& $0.178\pm0.001\;(0.029)$
& $0.136\pm0.002\;(0.046)$
& $\cdots$ \\
2 & 319 & 1  & 8.181 & 0.681 & 1486.520 & 29.687
& $-0.244\pm0.003\;(0.038)$
& $\cdots$
& $0.262\pm0.001\;(0.015)$
& $0.221\pm0.002\;(0.030)$
& $\cdots$ \\
3 & 296 & 3 & 8.140 & 0.569 & 1485.391 & 27.085
& $-0.226\pm0.003\;(0.035)$
& $\cdots$
& $0.258\pm0.001\;(0.018)$
& $0.215\pm0.002\;(0.032)$
& $\cdots$ \\
4 & 240 & 7 &  8.173 & 0.424 & 1551.047 & 15.942
& $-0.111\pm0.002\;(0.029)$
& $\cdots$
& $0.194\pm0.002\;(0.022)$
& $0.144\pm0.003\;(0.039)$
& $\cdots$ \\
5 & 226 & 1 &  8.093 & 0.688 & 1412.070 & 31.822
& $-0.305\pm0.003\;(0.032)$
& $\cdots$
& $0.282\pm0.001\;(0.017)$
& $0.237\pm0.002\;(0.028)$
& $\cdots$ \\
6 & 224 & 7 &  8.047 & 0.685 & 1364.087 & 33.036
& $-0.321\pm0.002\;(0.022)$
& $\cdots$
& $0.286\pm0.001\;(0.015)$
& $0.242\pm0.003\;(0.032)$
& $\cdots$ \\
7 & 140 & 6 &  8.139 & 0.397 & 1607.214 & 11.861
& $0.044\pm0.003\;(0.025)$
& $\cdots$
& $0.135\pm0.001\;(0.010)$
& $0.082\pm0.004\;(0.037)$
& $\cdots$ \\
8 & 125 & 2 &  8.092 & 1.024 & 1176.851 & 57.223
& $-0.568\pm0.004\;(0.033)$
& $\cdots$
& $0.342\pm0.002\;(0.015)$
& $0.263\pm0.004\;(0.032)$
& $\cdots$ \\
9 & 105 & 1 &  8.280 & 0.510 & 1628.285 & 15.454
& $-0.079\pm0.003\;(0.026)$
& $\cdots$
& $0.171\pm0.001\;(0.011)$
& $0.123\pm0.004\;(0.031)$
& $\cdots$ \\
\enddata
\tablecomments{
The ten largest chemical families are listed in descending order of the number of member stars. $R$ and $Z$ are in kpc, while $L_Z$ and $J_Z$ are in $\mathrm{kpc\,km\,s^{-1}}$. Chemical abundances are reported as median values with their uncertainties. Abundance dispersions of individual stars in the family are presented in parentheses. The complete machine-readable catalog contains all 18 chemical abundances for each family.
}
\end{deluxetable*}

\subsection{Distribution and characterization of the chemical families}

\begin{figure*}[htbp]
		\centering        
        \includegraphics[width=1\linewidth]{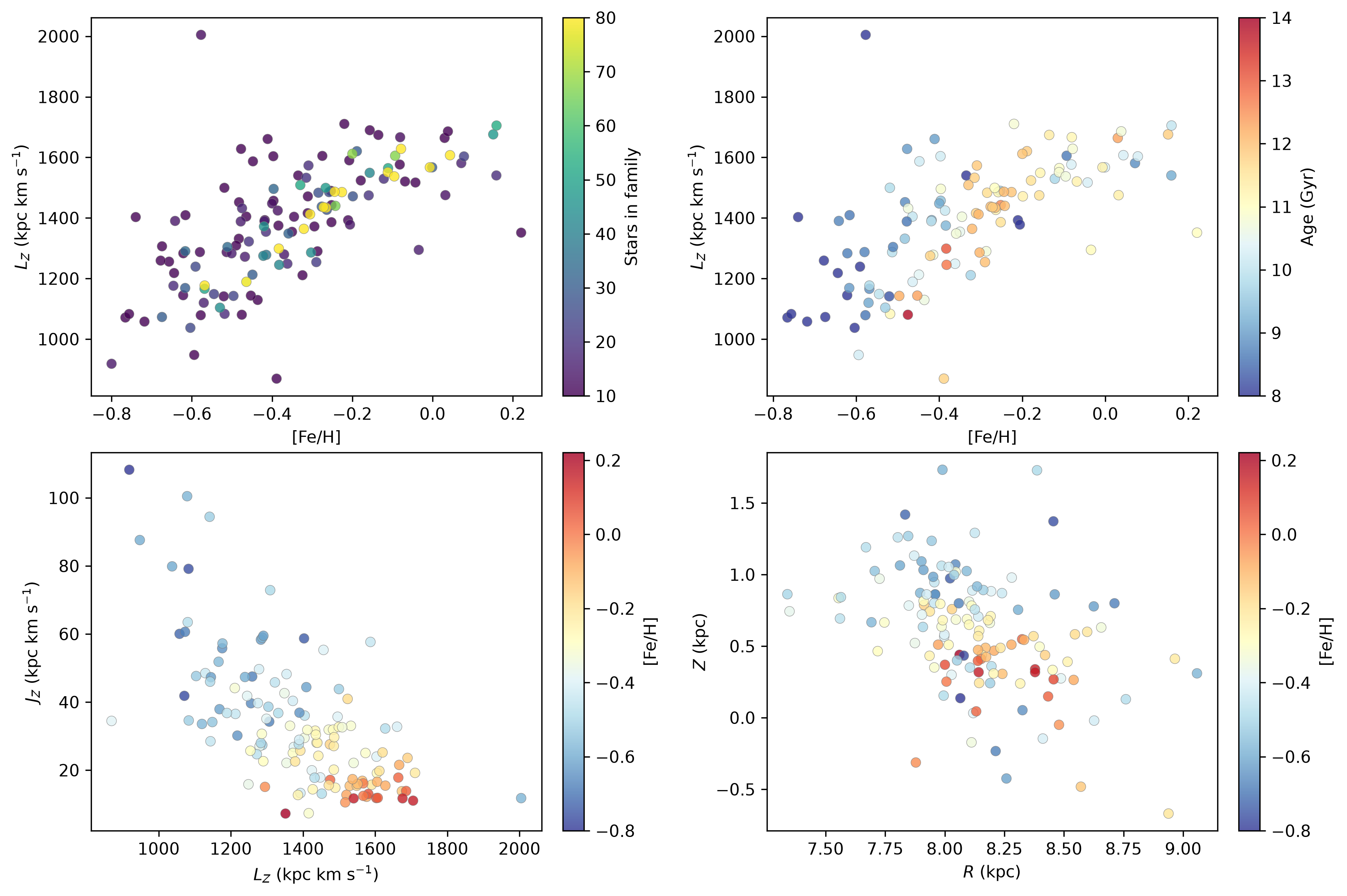}
		\caption{Distribution of the 133 identified chemical families in the chemical, orbital, and spatial parameter spaces.
        The upper-left panel shows the distribution of the families in the $[\mathrm{Fe/H}]$--$L_Z$ plane, where each point represents one chemical family and the color indicates the number of member stars in the family.
        The upper-right panel presents the same $[\mathrm{Fe/H}]$--$L_Z$ distribution, with the color scale representing the median age of each family.
        The lower-left panel shows the distribution of the families in the $L_Z$--$J_Z$ plane, color-coded by their median metallicity $[\mathrm{Fe/H}]$.
        The lower-right panel displays the spatial distribution of the chemical families in the Galactic cylindrical coordinate plane ($R$--$Z$), also color-coded by their median metallicity.
        }   
       \label{fig:r_z_lz_jz}
\end{figure*}

In this section, we investigate the spatial and orbital distributions of these chemical families.
Figure~\ref{fig:r_z_lz_jz} shows the distribution of the 133 identified chemical families in the [Fe/H]--$L_Z$, $L_Z$--$J_Z$, and $R$--$Z$ planes, where $L_Z$ ($J_\phi$) and $J_Z$ are the median orbital actions of the family in the azimuthal and vertical directions, respectively; $R$ and $Z$ are the families' median coordinates in the Galactic cylindrical coordinate system. The chemical families exhibit a broad distribution in these planes, suggesting that they represent ubiquitous populations of the high-$\alpha$ disk rather than products of peculiar events.

The positive [Fe/H]--$L_Z$ correlation is a well‑known feature of the high-$\alpha$ disk \citep[e.g.][]{2011ApJ...738..187L}, likely linked to the upside-down formation history as well as the dynamical heating history of the high-$\alpha$ disk, which results in larger velocity dispersion and scale height for older (more metal‑poor) stars \citep[e.g.][]{2025NatAs...9..101X}. Here we emphasize that, even at a fixed [Fe/H], we recover stellar families with different angular momenta, again demonstrating that clustering in the high‑dimensional parameter space is effective and does not simply rely on a strong correlation with [Fe/H]. In the $R$--$Z$ plane, the vast majority of the families are located at $Z>0$~kpc, which reflects a selection bias arising from the LAMOST survey volume.

Approximately 42\% (36,897/87,275) of our sample stars have age estimates from the catalog of \citet{2019MNRAS.484.5315W}, which provides ages for about 0.7 million red giant branch (RGB) stars derived from LAMOST spectra via a data-driven approach trained on asteroseismic age determinations. The upper-right panel of Figure~\ref{fig:r_z_lz_jz} presents the median age of member stars in each family across the [Fe/H]--$L_Z$ plane. It shows that, at a given [Fe/H], most families exhibit similar ages, though a few show notably distinct ages, suggesting inhomogeneous origins. However, the age estimates are subject to substantial systematic errors—in particular, the more metal-poor stars appear systematically younger, which contradicts the underlying trend established by precise and independent measurements of subgiant stars \citep{2022Natur.603..599X}. This discrepancy is likely due to imperfections in the age-training sample adopted by \citet{2019MNRAS.484.5315W}. Therefore, further investigation is still needed to robustly determine the age distribution of the families (Section~\ref{sec:age_metallicity}).

For each chemical family, we quantify the $L_Z$ dispersion of its constituent stars using the 16th–84th percentile width ($W_{L_{z}}=P_{84}-P_{16}$). Based on this dispersion, we classify the chemical families into two categories: an $L_Z$-dispersed category with $W_{L_{z}}\geq500$ $\mathrm{kpc\,km\,s^{-1}}$, and an $L_Z$-isolated category with $W_{L_{z}}<500$ $\mathrm{kpc\,km\,s^{-1}}$. In total, 103 families fall into the $L_Z$-dispersed category and 30 into the $L_Z$-isolated category.

\begin{figure*}[htbp]
		\centering
        \includegraphics[width=1\linewidth]{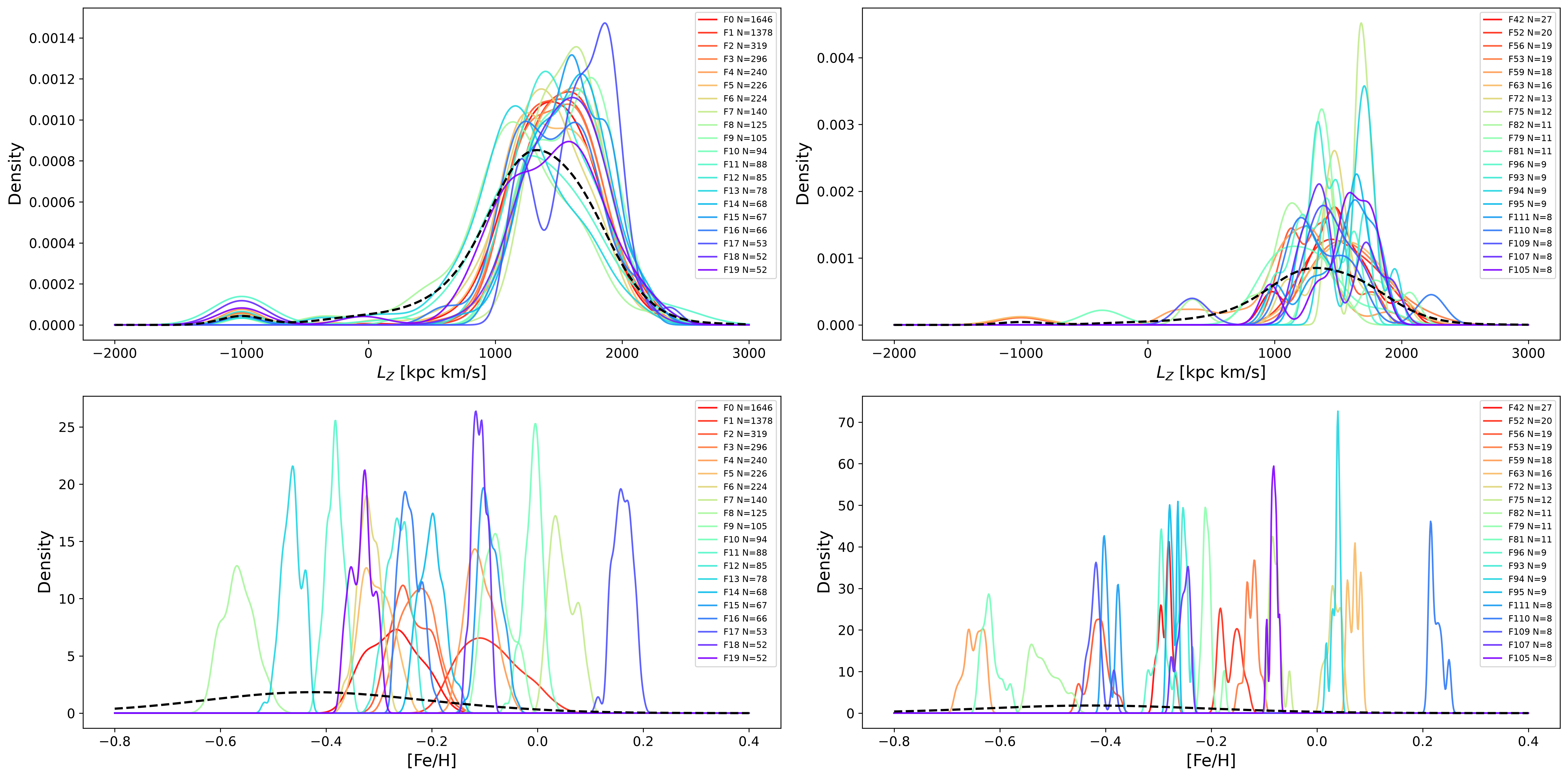}
		\caption{The distributions of angular momentum $L_Z$ (top panels) and metallicity [Fe/H] (bottom panels) for the identified chemical families. The left panels show non-isolated families (with $W_{L_{z}}\geq500$ $\mathrm{kpc\,km\,s^{-1}}$), while the right panels show isolated families (with $W_{L_{z}}<500$ $\mathrm{kpc\,km\,s^{-1}}$). Colors indicate individual families, sorted from largest (red) to smallest (purple). The dashed black curves represent the stars not assigned to any cluster. }
       \label{fig:lz_feh_kde}
\end{figure*}

Figure~\ref{fig:lz_feh_kde} presents the kernel density estimates of the $L_Z$ and [Fe/H] distributions for the 20 largest families in each category (the corresponding results for all families are shown in Figure~\ref{fig:lz_dis} and Figure~\ref{fig:feh_dis} in the Appendix). Most families in the $L_Z$-isolated category contain only a small number of stars ($<30$); whether their narrow $L_Z$ distribution reflects an intrinsic physical property or is simply a statistical artifact due to the small sample size remains unclear at present. Nevertheless, it is evident that nearly all families exhibit well-isolated [Fe/H] distributions, regardless of their $L_Z$ spread. This suggests that the member stars of each family likely formed within a short time interval, given the tight age–[Fe/H] relation of the high-$\alpha$ disk. For comparison, the figure also shows the distributions of the noise population — i.e., the stars not assigned to any cluster. These unassigned stars display broad distributions in both $L_Z$ and [Fe/H], in contrast to the chemically defined families.

We further examine the retrograde features ($L_Z<0$) in Figure~\ref{fig:lz_feh_kde} to determine whether they correspond to distinct chemically identified families. Among the 133 identified chemical families, 57 (42.86\%) contain at least one retrograde star, whereas only 10 (7.52\% ) have retrograde fractions exceeding 10\%. None of these families is dominated by retrograde stars, indicating that the retrograde features are associated with predominantly prograde chemical families rather than separately identified retrograde-dominated families. This is not unexpected, as our clustering is performed exclusively in chemical abundance space without imposing any kinematic constraints. Consequently, stars with similar chemical abundance patterns but different orbital properties may be assigned to the same family.
These retrograde stars may be related to the dynamically heated early Galactic disk population, often associated with the in-situ halo \citep{Bonaca_2017} or Splash population \citep{2019A&A...632A...4D, 2020MNRAS.497.1603G, 2020MNRAS.494.3880B, Sun_2026}, whose halo-like orbits may have resulted from ancient merger events.

\begin{figure*}[htbp]
		\centering
        \includegraphics[width=1\linewidth]{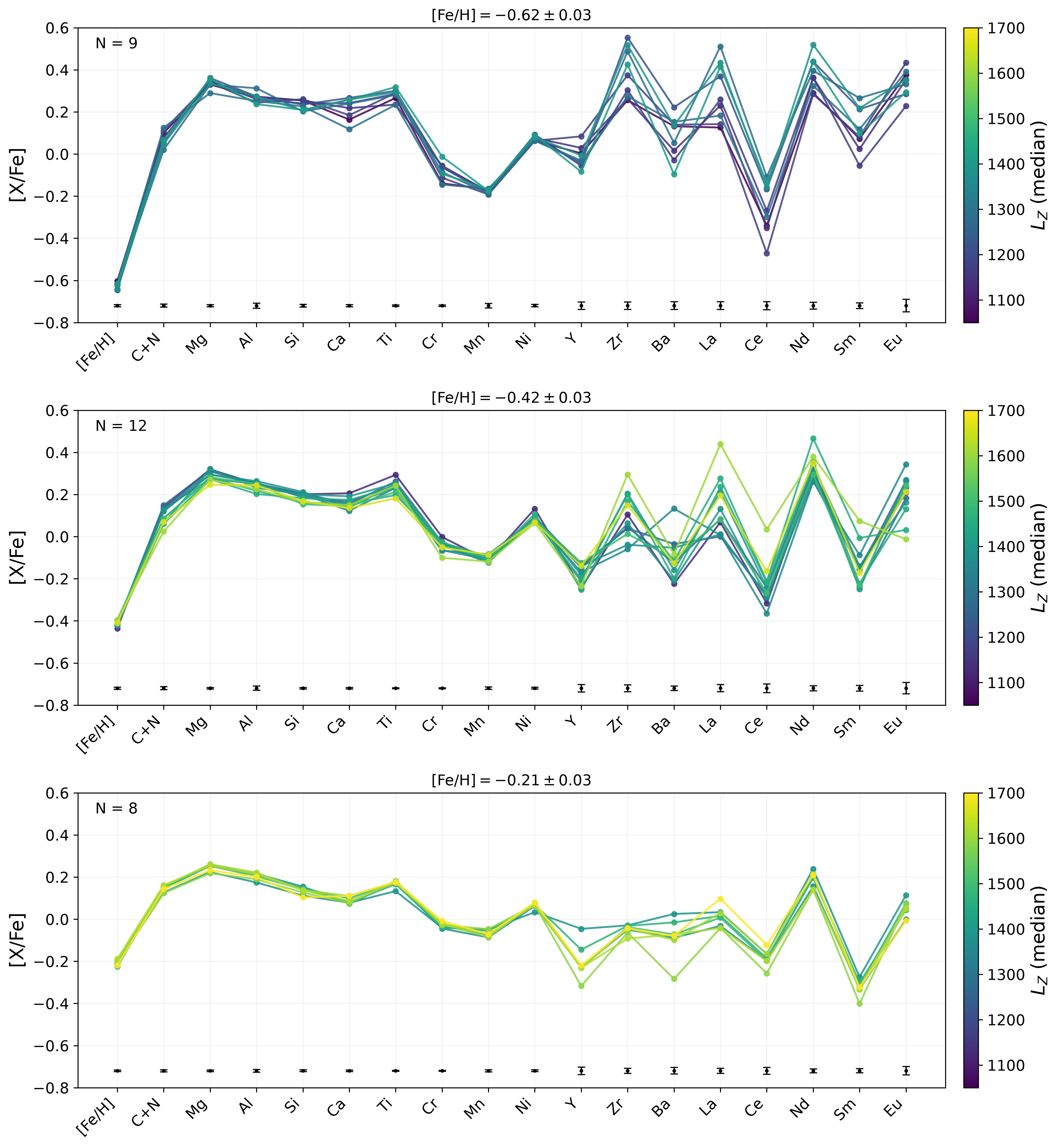}
		\caption{Median chemical abundance patterns of the identified families in three selected [Fe/H] bins. Each line represents the chemical pattern of each family and is color-coded by median angular momentum $L_Z$. 
        The x-axis shows the 18 elemental abundances adopted in this work, including light elements to neutron-capture species. The black error bars represent the typical uncertainties of the family median abundances within each metallicity bin. The number of identified families in each metallicity bin is indicated in the top-left corner of each panel.}
       \label{fig:lzfehpatterns}
\end{figure*}

Figure~\ref{fig:lzfehpatterns} shows the median abundance patterns of chemical families within three [Fe/H] bins, namely ${\rm [Fe/H]}=-0.62\pm0.03$, ${\rm [Fe/H]}=-0.42\pm0.03$, ${\rm [Fe/H]}=-0.21\pm0.03$. Within each [Fe/H] bin, different families exhibit distinguishable abundance patterns for a wide range of elements. The difference in element abundance is particularly large for neutron-capture elements, while that for $\alpha$-elements (Mg, Si, Ca, and Ti) and iron-peak elements (Cr, Mn, Ni) are smaller but still noticeable. These results verify previous findings that neutron-capture elements show larger intrinsic scatter among disk stellar populations \citep[e.g.][]{2024ApJ...972...69M, 2026MNRAS.tmp.1588W}. To highlight these variations, the relative abundance patterns after subtracting the median value of each element are shown in Figure~\ref{fig:patternsdelta} in the Appendix.

\section{Discussion}
\label{sec:discussion}

\subsection{Feasibility of low‑resolution spectroscopy for chemical tagging}
Our results demonstrate that abundances derived from LAMOST low‑resolution ($R\lesssim1800$) spectra are capable of identifying chemically coherent groups in high‑dimensional abundance space, provided that a sufficiently large number of elements—including neutron‑capture species—are included. This confirms that even with larger measurement uncertainties (0.1–0.2 dex) for neutron‑capture elements, the combined information from these elemental abundances provides useful discriminating power \citep{ 2022ApJ...927..209T, 2024ApJ...972...69M, 2026MNRAS.tmp.1588W}. 

The most plausible explanation for this success is the astrophysical correlations among different elemental abundances. Using APOGEE data, \citet{2022ApJ...927..209T} quantitatively investigated correlations among elemental abundances and demonstrated that these correlations contain important information for chemical tagging beyond that provided by individual abundance measurements. Our results are consistent with this picture. In particular, the chemically homogeneous mock sample, in which both intrinsic abundance variations and astrophysical correlations among elements are removed, produces predominantly a single cluster or no clusters across the explored DBSCAN parameter combinations. This suggests that measurement uncertainties alone do not systematically generate the multiple chemical structures identified in the observed sample, although the mock test does not isolate the contribution of elemental correlations from that of intrinsic abundance variations.

Given the current measurement uncertainties, coherent multi-element abundance patterns therefore provide a potentially useful means of identifying chemical structures in low-resolution spectroscopic data. In the future, more precise abundance measurements may enable independent tests of the identified families through measurements of their intrinsic abundance dispersions.

\subsection{Chemical Diversity in the High-$\alpha$ Disk}
The identification of 133 chemical families with distinct multi-element abundance patterns reveals chemically distinguishable structures within the Galactic high-$\alpha$ disk. Although these families account for only a subset of the stellar sample, their systematic abundance differences at nearly identical [Fe/H] demonstrate that the identified structures capture chemical variations beyond the global metallicity sequence. These variations are particularly prominent among neutron-capture elements, suggesting that multi-element abundance patterns can provide additional information about the chemical enrichment history of the high-$\alpha$ disk.

The observed differences may reflect variations in the relative contributions of different nucleosynthetic sources, including supernovae and asymptotic giant branch stars, or differences in the chemical enrichment histories of the gas from which the stars formed. The current study identifies only $\sim5$\% of the stars as belonging to chemically distinguishable families. To this end, the high-$\alpha$ disk appears largely chemically smooth at the precision of the current abundance estimates. However, given the relatively large uncertainties in the measured elemental abundances, it remains unclear whether a substantially larger fraction of the high-$\alpha$ disk population exhibits distinct abundance patterns that are unresolved in the current data. More precise abundance measurements from future high-resolution spectroscopic surveys, such as 4MOST \citep{2019Msngr.175....3D, 2019Msngr.175...35B}, will provide a more comprehensive picture of the chemical diversity within the high-$\alpha$ disk.

\subsection{Linking chemical families to the tight stellar age–metallicity relation}\label{sec:age_metallicity}
A puzzle arises when we consider the well‑established age–metallicity relation of the high‑$\alpha$ disk. \citet{2022Natur.603..599X} showed that at a given [Fe/H], the intrinsic age scatter is only $\sim$0.8~Gyr, despite the overall age range spanning $\sim$5~Gyr. An important question is how the different chemical abundance patterns identified in this work relate to the tight age--metallicity relation of the high-$\alpha$ disk. Do these patterns trace populations with different stellar ages, or can similar chemical abundance patterns arise over extended periods of enrichment? Understanding these questions requires a careful characterization of the age distributions of the chemical families. Unfortunately, current age estimates for LAMOST red giants derived from spectra using data‑driven methods suffer from relatively large uncertainties, including 10–30\% random errors and substantial systematic errors for relatively metal-poor, high-$\alpha$ disk stars \citep[][]{2017ApJ...841...40H, 2019MNRAS.484.5315W, 2023A&A...675A..26W}. Additionally, spectroscopically data-driven age may exhibit artificial correlations with the abundances themselves, which would bias any comparison. Independent age determinations from asteroseismology, available for a growing number of field giants (e.g., from Kepler \citep{2025ApJS..276...69P, 2023MNRAS.520.1913W} and TESS \citep{2015JATIS...1a4003R}), provide a promising route, but the sample of high‑$\alpha$ disk stars with such data is limited. Future missions such as PLATO \citep{2017AN....338..644M,2025ExA....59...26R} and the Earth 2.0 mission \citep[ET][]{2022arXiv220606693G, 2026SPIE14145E..09G} are expected to deliver asteroseismic ages for large populations of old disk stars, which would directly characterize the age distribution within each chemical family.

In the meantime, a complementary approach is to focus on subgiant stars, for which isochrone ages can be derived with much higher precision \citep{2022Natur.603..599X} than for giants. By applying our clustering analysis to a sample of high‑$\alpha$ subgiants with precise ages, one could assess whether the identified families indeed occupy narrow age ranges and whether the family sequence matches the age–metallicity relation.

\section{Conclusion}
\label{sec:conclusion}
Chemical substructures provide crucial clues to the formation history of the Galactic high‑$\alpha$ disk. In this work, we apply the DBSCAN clustering method to a sample of 87,275 high‑$\alpha$ red giants, using 18-dimensional elemental abundances derived from LAMOST low‑resolution spectra. Through validation using common stars with APOGEE and using mock data set, we demonstrate that despite their larger individual uncertainties, abundances from low‑resolution spectroscopy collectively offer essential discriminating power in high‑dimensional abundance space and enable the identification of chemically coherent stellar families. The discriminating power is largely encoded in the astrophysical correlation among different elemental abundances.

After scanning the parameter space of the DBSCAN procedure and merging the results from different parameter combinations, we identify 133 candidate chemical families in the Galactic high-$\alpha$ disk, comprising 4642 unique stars (5.3\% of the sample). These families exhibit diverse multi-element abundance patterns beyond differences in their overall metallicities: at nearly identical [Fe/H], they show systematic differences in multiple abundance ratios, most prominently among neutron-capture elements.
These results reveal chemical diversity beyond the global metallicity trend and suggest that the identified families may preserve information about variations in nucleosynthetic contributions and chemical enrichment histories. The resulting catalog provides a basis for future investigations of the chemical evolution of the high-$\alpha$ disk. 

\section{Acknowledgements}
We thank the anonymous referee for valuable comments and suggestions that helped improve the manuscript.

We acknowledge financial support from the National Natural Science Foundation of China (NSFC; Grant No. 12588202, No. 2022000083) and the National Key R\&D Program of China Grant (No.2022YFF0504200). M.Z. acknowledge financial support from NSFC No. 12303025.

This work made use of the data from LAMOST (Large Sky Area Multi-Object Fiber Spectroscopic Telescope, also known as the Guoshoujing Telescope) (https://cstr.cn/31118.02.LAMOST). LAMOST is a Chinese national mega-science facility, operated by National Astronomical Observatories, Chinese Academy of Sciences. Funding for the project has been provided by the National Development and Reform Commission. LAMOST is operated and managed by the National Astronomical Observatories, Chinese Academy of Sciences. The LAMOST website is https://www.lamost.org. 


\bibliography{sample631}{}
\bibliographystyle{aasjournal}


\appendix

\section{$L_Z$ and [Fe/H] distribution for the 133 chemically coherent families in high-$\alpha$ disk}
\begin{figure*}[htbp]
		\centering
        \includegraphics[width=1\linewidth]{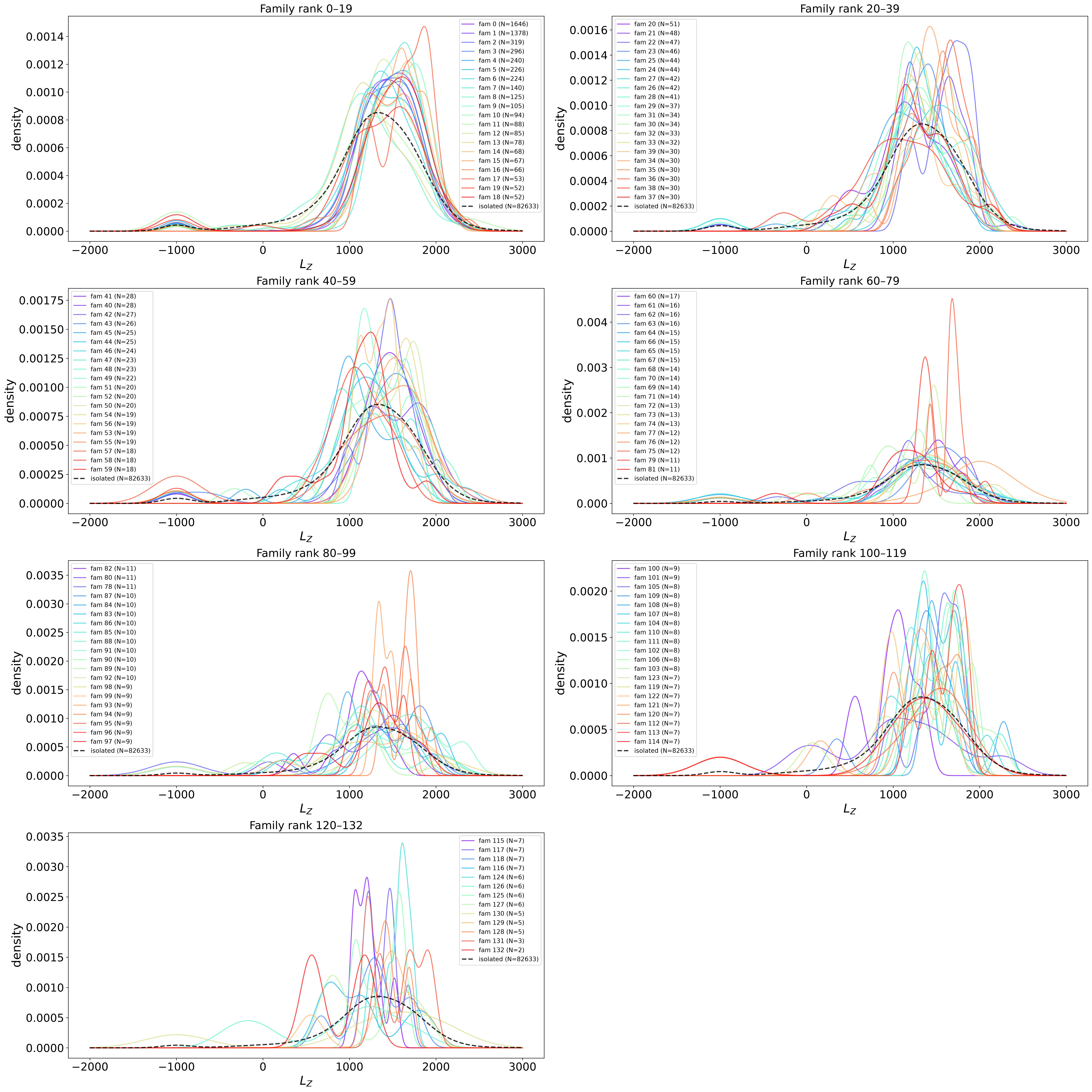}
		\caption{Distribution of the angular momentum $L_z$ for the identified chemical families in the high-dimensional abundance space. Each colored curve represents the kernel density distribution of $L_z$ for an individual chemical family after the hierarchical merging procedure. Different panels display different subsets of chemical families for clarity. Many families exhibit relatively compact and distinct $L_z$ distributions, suggesting that chemically coherent structures identified in abundance space may also retain characteristic dynamical properties. Several families show broad or multi-peaked $L_z$ distributions, indicating possible dynamical mixing, radial migration, or the presence of chemically similar but dynamically complex populations.}
       \label{fig:lz_dis}
\end{figure*}

\begin{figure*}[htbp]
		\centering
        \includegraphics[width=1\linewidth]{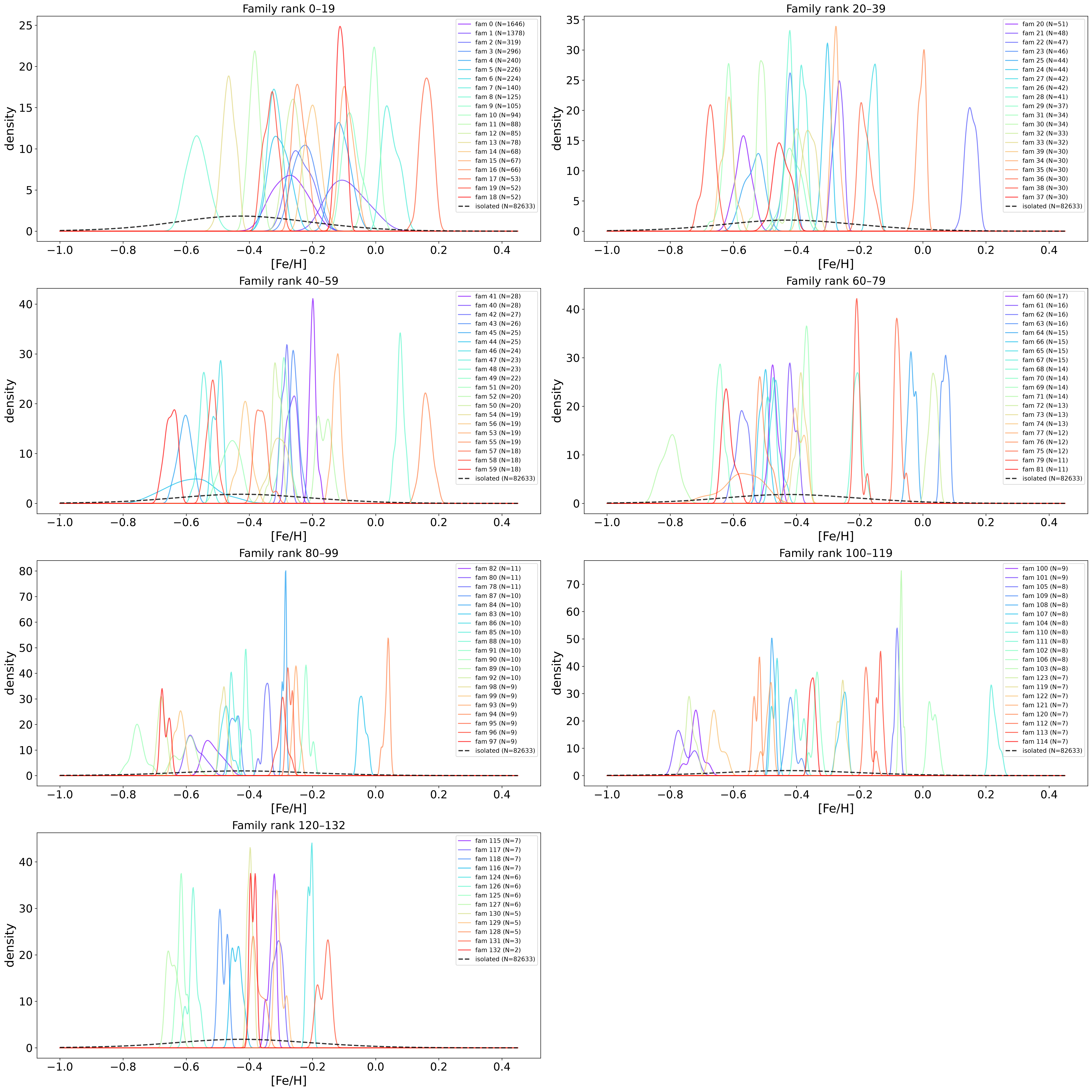}
		\caption{Same as Figure~\ref{fig:lz_dis}, but for the metallicity distribution ([Fe/H]) of the identified chemical families.}
       \label{fig:feh_dis}
\end{figure*}

Figure~\ref{fig:lz_feh_kde} in the main text showed the $L_Z$ and [Fe/H] distribution for only a fraction of the high-$\alpha$ disk families. The results for the full sample of the 133 families are shown in Figure~\ref{fig:lz_dis} and Figure~\ref{fig:feh_dis}. 

\begin{figure*}[htbp]
		\centering
        \includegraphics[width=1\linewidth]{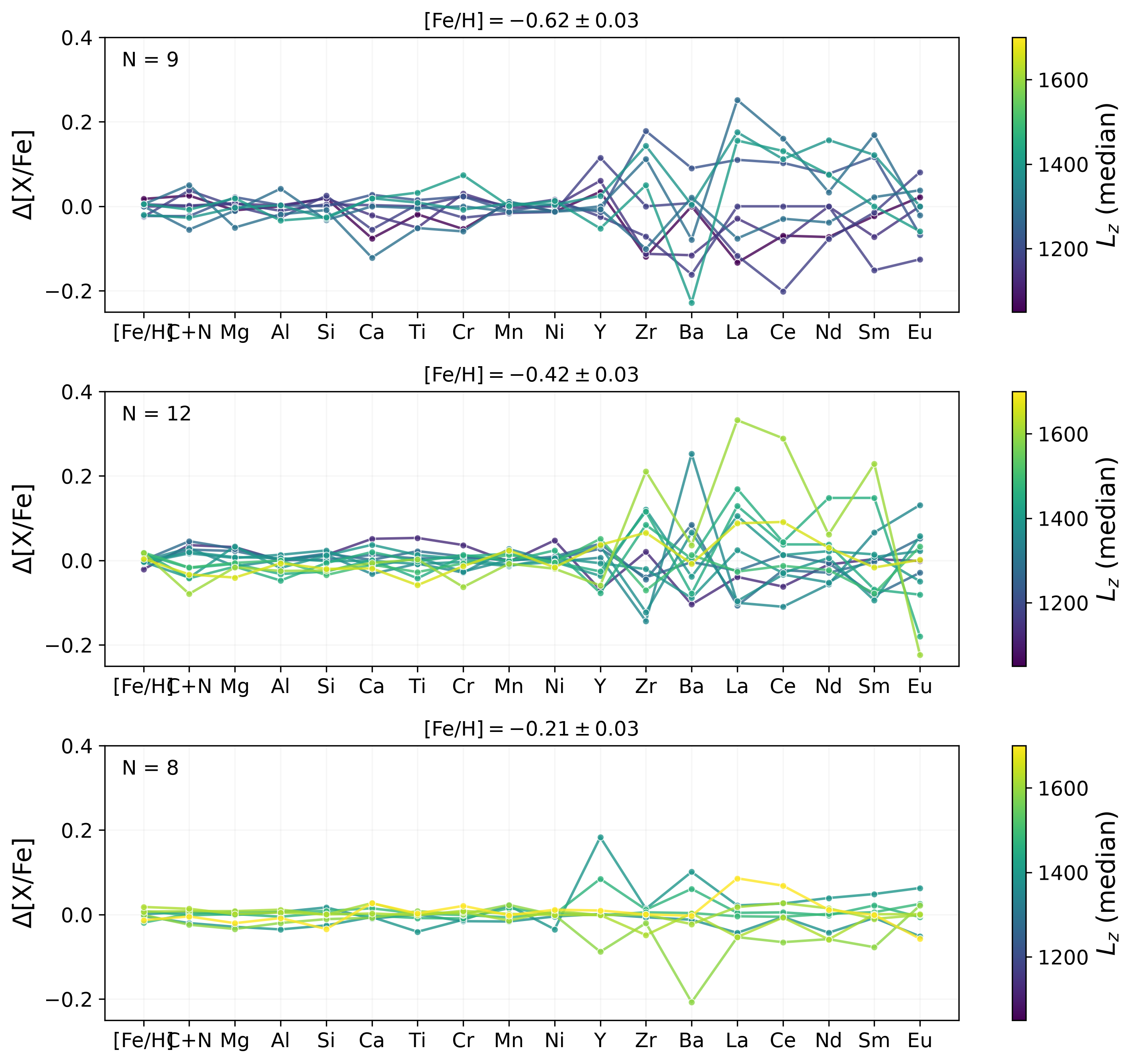}
		\caption{Same as Figure~\ref{fig:lzfehpatterns}, but showing $\Delta [X/{\rm Fe}]$, i.e., the abundance of each element relative to the median of all families in the same [Fe/H] bin. This highlights the relative contribution of individual elements to the chemical separation among families. }
       \label{fig:patternsdelta}
\end{figure*}

Figure~\ref{fig:patternsdelta} shows the relative abundance patterns for families selected for three [Fe/H] windows, which is similar to Figure~\ref{fig:lzfehpatterns} but after subtracting the median value of [X/Fe] of individual families for each element X.  

\end{document}